\documentclass[letterpaper, 12pt]{article}
\usepackage[margin=1in]{geometry}

\usepackage{generic}

\usepackage{amsmath}

\allowdisplaybreaks
\usepackage{xcolor}
\usepackage{footnote}
\usepackage{algorithm}
\usepackage{algpseudocode}
\usepackage{algorithmicx}
\usepackage{multirow} 
\usepackage{booktabs}
\usepackage{graphicx}
\usepackage{amssymb}
\usepackage{amsbsy}

\usepackage{amsthm}

\usepackage{array}
\usepackage{longtable}
\usepackage{epstopdf}
\usepackage{pbox}
\usepackage{breqn}
\usepackage{mathrsfs}
\usepackage{multicol}
\usepackage{supertabular}
\usepackage{enumerate}
\usepackage[colorinlistoftodos]{todonotes}
\usepackage{url}
\usepackage[justification=centering]{caption}
\usepackage{tabu}
\usepackage{subcaption}
\usepackage{graphics} 
\usepackage{epsfig} 

\usepackage{mathtools}

\usepackage{tikz}
\usepackage{tikz-network}

\usepackage{cases}
\usepackage{bm}
\usepackage{cite}

\usepackage{aligned-overset}

\newtheorem{thm}{Theorem}
\newtheorem*{thm*}{Theorem}
\newtheorem{prp}{Proposition}

\newtheorem{lmm}{Lemma}
\newtheorem{asm}{Assumption}

\newtheorem{cor}{Corollary}

\theoremstyle{definition}

\newtheorem{dfn}{Definition}

\newcommand{\thistheoremname}{}
\newtheorem*{genericthm*}{\thistheoremname}
\newenvironment{namedthm*}[1]
  {\renewcommand{\thistheoremname}{#1}%
  \begin{genericthm*}}
  {\end{genericthm*}}

\title{\LARGE
Resilient Control of Dynamic Flow Networks Subject to Stochastic Cyber-Physical Disruptions
}

\author{Yu Tang and Li Jin
\thanks{This work was in part supported by US NSF Award CMMI-1949710, USDOT Award \# 69A3551747124 via the C2SMART Center, NYU Tandon School of Engineering, SJTU UM Joint Institute, and J. Wu \& J. Sun Endowment Fund.}
\thanks{Y. Tang is with the Tandon School of Engineering, New York University, USA. L. Jin is with the UM Joint Institute and with the Department of Automation, Shanghai Jiao Tong University, China (emails: tangyu@nyu.edu, li.jin@sjtu.edu.cn).}
}

\begin{document}
\maketitle

\begin{abstract}
Modern network systems, such as transportation and communication systems, are prone to cyber-physical disruptions and thus suffer efficiency loss.
This paper studies network resiliency, in terms of throughput, and develops resilient control to improve throughput. We consider single-commodity networks that admit congestion propagation.
We also apply a Markov process to model disruption switches.
For throughput analysis, we first use insights into congestion spillback to propose novel Lyapunov functions 
and then exploit monotone network dynamics to reduce computational costs of verifying stability conditions. For control design, we show that
(i) for a network with infinite link storage space, there exists an open-loop control that attains the min-expected-cut capacity;
(ii) for a network with observable disruptions that restrict maximum sending and/or receiving flows, there exists a mode-dependent control that attains the expected-min-cut capacity; (iii) for general networks, there exists a closed-loop control with throughput guarantees. We also derive lower bounds of resiliency scores for a set of numerical examples and verify resiliency improvement with our method.
\end{abstract}


\section{Introduction}

\subsection{Motivation}
Dynamic flow networks are widely used to model engineering systems including transportation systems \cite{schmitt2018exact} and communication networks \cite{mandjes2003models}.
These systems are susceptible to disruptions both in physical and cyber parts.
In the physical part, link flows can be disrupted by capacity-reducing events such as traffic incidents \cite{jin2018stability}.
In the cyber part, unreliable state observation or faulty feedback actuation may occur, which degrades the effectiveness of feedback controllers and lead to physical losses \cite{sandberg2015cyberphysical}. Typically, both types of disruptions are hard to predict and thus need treatment in a stochastic manner. For instance, probabilistic models were used to evaluate freeway capacity, which is supported by field studies \cite{polus2002stochastic}. This modeling technique is also applicable to sensing faults \cite{dong2012fault}. Since network systems involve very large numbers of cyber-physical components, it is neither economically feasible nor technically necessary to prevent all disruptions.
Instead, a practical solution is to integrate disruptions into modeling control schemes \cite{cardenas2009challenges}. However, limited studies have discussed such method for dynamic flow networks subject to stochastic disruptions.

In this paper,  we first evaluate network throughput as a measure of resiliency against stochastic cyber-physical disruptions, arising from reliability failures \footnote{In this paper, reliability failures refer to temporary malfunction of components of controlled dynamic flow networks, such as node/link breakdown and sensing/actuation faults. They do not include communication delay or malicious cyber attacks. Though this paper mainly considers sensing failures for cyber disruptions, our approach can be applied to address actuation faults with minor modifications.}, and then design control strategies that mitigate throughput losses. To this end, we consider single-commodity networks that have found applications in real systems, such as freeway systems \cite{schmitt2018exact}. Although single-commodity networks could simplify real-word cases, they are worth studying since multi-commodity networks require origin-destination (OD) demands that are hard to acquire in practice. We use a finite-state Markov process to model the occurrence and clearance of disruptions.
Then we investigate stability condition of networks, which allows for resiliency quantification and resilient control configuration. Particularly, our discussion on control design, inspired by the classical max-flow min-cut theorem \cite{dantzig1955max} for static flow networks, proposes \emph{min-expected-cut capacity} (MECC) and \emph{expected-min-cut capacity} (EMCC) and reveals their relation to throughput, network storage space, disruptions and control laws. In case of a general network whose throughput is not guaranteed with these capacities, we present a closed-loop control with lower-bounded throughput.

\subsection{Related work}
Previous work on dynamic flow networks typically considered a nominal or robust setting. In the disruption-free case, in-depth stability analysis was provided for networks characterized by various flow structures including monotone dynamics, nonlinear demand/supply constraints, and congestion spillback \cite{como2014throughput,coogan15,nilsson2021strong,massai2020stability,coogan2016stability}. Besides, optimal routing control has been extensively investigated for dynamical flow networks. \cite{como2014throughput,jafari2019structural}. Robust control strategies, along with resiliency analysis, were developed in the face of physical disruptions \cite{johansson2018optimization,fan2004robustness,alpcan2008robust,como13i,como2017resilient,savla2020network}. 
However, the above work assumes perfect sensing and actuation; the resiliency against cyber disruptions remains unclear. Besides, robust control handles disturbances with uncertainty sets \cite{savla2020network}; it does not apply to recurrently switching disruptions. 

The model we consider belongs to a subclass of piecewise-deterministic Markov processes (PDMP) \cite{teel2014stability}, sometimes known as Markov jump nonlinear systems (MJNS) \cite{shi2015survey}, where continuous states (traffic densities) evolve according to a set of ordinary differential equations and a discrete state (disruption mode) determines the mode of the continuous dynamics \cite{davis84}.
Such a formulation allows analysis of joint impacts of cyber and physical disruptions on network resiliency. Although the general theories of PDMPs \cite{davis84,benaim15,mesquita2010construction} and MJNSs \cite{de2006robust,chatterjee2007stability} have been extensively investigated,  the implementation is still challenging due to nonlinear flow dynamics. 
Typical Lyapunov functions, such as piecewise quadratic functions \cite{chatterjee2007stability}, fail to capture congestion dynamics and thus only lead to trivial stability analysis; see our discussion in Section~\ref{sec_analysis}. It is vital to design appropriate Lyapunov functions for analyzing our model.

Our work is also related to stochastic fluid models \cite{dai95ii,chen1992fluid}. These models are applied to performance analysis or admission/priority control for servers subject to demand or service disruptions \cite{yu2004perturbation,wardi2009unified}. 
Currently, limited results consider congestion spillback over stochastic flow networks, let alone the corresponding control design. Only a few results were developed for networks with special structures, such as parallel links \cite{jin2018stability, xie2020resilience} and serial links \cite{sun2004perturbation,jin2018analysis}. 
To our best knowledge, the general stochastic flow networks are still not well investigated. 

\subsection{Our contributions}
This paper focuses on the two following questions:
\begin{enumerate}[(i)]
    \item How to quantify resiliency against stochastic disruptions, especially for networks with congestion propagation? 
    \item How to attain resiliency-by-design or improve resiliency? 
\end{enumerate}

We define \emph{resiliency score} as a ratio of disrupted network throughput to its nominal value. Here the disrupted  (resp. nominal) throughput means the maximal inflow under which the network with (resp. without) disruptions can be stabilized, i.e. traffic densities in all links being bounded on average. The max-flow min-cut theorem states that the nominal throughput equals \emph{min-cut capacity} \cite{dantzig1955max}, but we can hardly resolve the disrupted throughput in an analytical way. It is because our PDMP model allows complicated disruptions, including the physical ones creating new bottlenecks and the cyber ones inducing mismatches between control instructions and actual inter-link flows. Thus we address (i) by deriving and sharpening lower bounds of throughput. The lower bounds are obtained from a set of stability conditions built on the Foster-Lyapunov criterion \cite{meyn93}. To formulate the stability conditions, we use the insights into network-wide congestion propagation to propose a set of novel Lyapunov functions. We also exploit monotone network dynamics to simplify the condition verification from over unbounded sets to over only compact sets, which saves computational costs. 

As indicated above, the max-flow min-cut theorem could be compromised. Thus we consider its variants for resiliency-by-design control. Meanwhile, a more practical concern is that full observation of disruption modes and network states (traffic densities) could not be available to control design.
To answer (ii), we first show that there exists an open-loop control that attains the MECC if all links have infinite storage space. Second, we show that there exists a mode-dependent control that attains the EMCC if the disruptions restrict maximum sending/receiving flows. The above results resemble the classical max-flow min-cut theorem. Third, we propose a density-dependent control, that is throughput-guaranteed, for general networks disrupted stochastically. Finally, we use numerical examples to demonstrate that our methods can enhance network resiliency.

The rest of this paper is organized as follows.
Section~\ref{sec_model} introduces our PDMP model for networks subject to stochastic cyber-physical disruptions.
Section~\ref{sec_analysis} analyzes the resiliency of this network.
Section~\ref{sec_control} presents resilient control design.
Section~\ref{sec_conclude} summarizes the main conclusions and discusses future directions. 
\section{Dynamic flow network with cyber-physical disruptions}
\label{sec_model}

Consider a single-origin-single-destination directed network $\mathcal{G}=(\mathcal{V},\mathcal{E})$, where $\mathcal{V}$ and $\mathcal{E}$ denote the node set and the link set, respectively. Though single-commodity networks may have multiple origins and destinations, we can introduce one artificial origin and destination so that we obtain single-origin-single-destination networks where demands are routed from the artificial origin to real ones by proportions. For ease of presentation, we assume that the network is acyclic. Note that the proposed method can be applied to cyclic networks as well; see our discussion in Appendix~\ref{app_cyclic}. We denote the starting and ending nodes of link $e$ by $\sigma_e$ and $\tau_e$, respectively. We number the origin node as $v_o$ and the destination node as $v_d$. The origin is subject to a constant inflow of $\alpha\in\mathbb R_{\ge0}$. Without loss of generality, we assume that the flow enters the network via link $e_o$; see Fig.~\ref{fig_1}.
\begin{figure}[htbp]
    \centering
    \resizebox{0.5\linewidth}{!}{
    \begin{tikzpicture}
        \Vertex[x=-1.3,style={color=white}]{A}
        \Vertex[label=$v_o$,x=1,fontscale=1.4]{B}
        \Vertex[x=2.5,y=1.8]{C}
        \Vertex[x=2.5,y=-1.8]{E}
        \Vertex[label=$\sigma_e$,x=4,y=0,fontscale=1.4]{G}
        \Vertex[label=$\tau_e$,x=5.5,y=0,fontscale=1.4]{I}
        \Vertex[x=7,y=1.8]{K}
        \Vertex[x=7,y=-1.8]{M}
        \Vertex[label=$v_d$,x=8.5,y=0,fontscale=1.4]{N}
        
        \Vertex[x=4.1,y=1,size=0.5,style={white}]{G_up}
        \Vertex[x=4.5,y=1.1,size=0.5,style={white},label=$\mathcal{E}_e^-/\mathcal{E}_{\sigma_e}^-$,fontscale=1.4]{G_up_annotation}
        \Vertex[x=4.1,y=-1,size=0.5,style={white}]{G_down}
        \Edge[style={dashed},bend=60](G_down)(G_up)
        
        \Vertex[x=5.4,y=1,size=0.5,style={white}]{I_up}
        \Vertex[x=5.4,y=-1,size=0.5,style={white}]{I_down}
        \Vertex[x=5.1,y=-1.1,size=0.5,style={white},label=$\mathcal{E}_e^+/\mathcal{E}_{\tau_e}^+$,fontscale=1.4]{I_down_annotation}
        \Edge[style={dashed},bend=60](I_up)(I_down)
        
        \draw[draw=black,dashed] (-0.8, -1.1) rectangle ++(4.1, 2.2);
        \Vertex[x=0,y=1.1,size=0.5,style={white},label=$\mathcal{A}_e^-$,fontscale=1.4]{A_set_up}
        
        \draw[draw=black,dashed] (6.2, -1.1) rectangle ++(4.1, 2.2);
        \Vertex[x=9.5,y=1.1,size=0.5,style={white},label=$\mathcal{A}_e^+$,fontscale=1.4]{A_set_down}
        
        
        \Edge[Direct,label=$e_o$,fontscale=1.4](A)(B)
        \Edge[Direct,label=$\cdots$](B)(C)
        \Edge[Direct,label=$\cdots$](B)(E)
        \Edge[Direct,
        fontscale=1.4](C)(G)
        \Edge[Direct,
        fontscale=1.4](E)(G)
        \Edge[Direct,
        fontscale=1.4](I)(K)
        \Edge[Direct,
        fontscale=1.4](I)(M)

        \Edge[Direct,label=$\cdots$](M)(K)
        \Edge[Direct,label=$\cdots$](C)(E)
        \Edge[Direct,label=$\cdots$](C)(K)
        \Edge[Direct,label=$\cdots$](E)(M)
        \Edge[Direct,label=$\cdots$](K)(N)
        \Edge[Direct,label=$\cdots$](M)(N)
        \Edge[Direct,label=$e$, position={below=1mm},fontscale=1.4](G)(I)

        \Text[x=0.5,y=-0.8]{Origin}
        \Text[x=9.2,y=-0.8]{Destination}
        
    \end{tikzpicture}
    }
    \caption{A single-origin-single-destination network: we denote by $\mathcal{E}_e^-$ (resp. $\mathcal{E}_{\sigma_e}^-$) the set of incoming links of link $e$ (resp. node $\sigma_e$), by $\mathcal{E}_e^+$ (resp. $\mathcal{E}_{\tau_e}^+$) the set of outgoing links of link $e$ (resp. node $\tau_e$), by $\mathcal{A}_e^-$ the set of links upstream of link $e$, and by $\mathcal{A}_e^+$ the set of links downstream of link $e$.}
    \label{fig_1}
\end{figure}
Following the convention \cite{como13i}, we consider  traffic density (mass per unit) as the network state and denote by $X_e(t)$ the density of link $e$ at time $t$. We also assume that all links have the same unit length for convenience of computing densities.  
For link $e$ with finite storage space, $X_e(t)$ can only take values from a closed interval $[0,x_e^{\max}]$, where $x_e^{\max}<\infty$ is called the \emph{jam density}. 
For link $e$ with infinite storage space, $X_e(t)$ can take values from $\mathbb R_{\ge0}$ and we let $x_e^{\max}=\infty$. In particular, we assume that link $e_o$ has infinite storage space; this ensures that no traffic is rejected into $e_o$. We use $\mathcal X_e$ to denote the set of $X_e(t)$, $\mathcal{X}=\prod_{e\in\mathcal{E}}\mathcal X_e\subseteq\mathbb{R}^{\mathcal{E}}_{\geq0}$ to denote the set of the state vector $X(t)$, where $\mathbb{R}^{\mathcal{E}}_{\geq0}$ stands for a set of non-negative vectors whose components are
indexed by elements of $\mathcal{E}$.

For a node $v\in\mathcal{V}$, we let $\mathcal{E}_v^-:=\{e\in\mathcal{E}|\tau_e=v\}$ (resp. $\mathcal{E}_v^+:=\{e\in\mathcal{E}|\sigma_e=v\}$) denote the set of its incoming (resp. outgoing) links; for a link $e\in\mathcal{E}$, we use $\mathcal{E}^-_e:=\{i\in\mathcal{E}|\tau_i=\sigma_e\}$ (resp. $\mathcal{E}^+_e:=\{j\in\mathcal{E}|\sigma_j=\tau_e\}$) to denote the set of its upstream (resp. downstream) adjacent links. 
Clearly, we have $\mathcal{E}_v^-=\mathcal{E}_e^-$ if $v=\sigma_e$ and $\mathcal{E}_v^+=\mathcal{E}_e^+$ if $v=\tau_e$. 
We also let $\mathcal{P}:=\{(e,j)|e\in\mathcal{E},j\in\mathcal{E}^+_e\}$ denote the set of ordered pairs of adjacent links. It follows $|\mathcal{P}|=\sum_{e\in\mathcal{E}}|\mathcal{E}^+_e|$, where $|\cdot|$ denotes the cardinality of a set. 
We say that link $e$ is accessible from link $i$, denoted by $i\to e$, if there exists a directed path starts with link $i$ and ends at link $e$. For link $e\in\mathcal E$, let $\mathcal{A}^-_e:=\{i\in\mathcal{E}|i\to e\}$ be the set of links from which link $e$ is accessible, and let $\mathcal{A}^+_e:=\{j\in\mathcal{E}|e\to j\}$ be the set of links that are accessible from link $e$. 

In the rest of this section, we first define flow functions, control laws, and disruption modes (Section~\ref{sub_flow}), and then specify the network's dynamics as a piecewise-deterministic Markov process (Section~\ref{sub_pdmp}). Finally, we define network stability and resiliency score (Section~\ref{sub_add}).

\subsection{Flows, control laws and disruptions}
\label{sub_flow}

Below we introduce the essential definitions and assumptions for our network model. The illustrative examples are also provided.

\subsubsection{Sending/receiving flows}
The \emph{sending flow} of link $e$ is specified by $f_e:\mathcal X_e\to\mathbb R_{\ge0}$. It stands for the maximum outflow from link $e$ given a state $x_e$ The \emph{receiving flow} of link $e$ is specified by $r_e:\mathcal X_e\to\mathbb R_{\ge0}$. It is recognized as the maximum inflow allowed into link $e$ given a state $x_e$. We assume that these flows satisfy:
\begin{asm}[Sending/receiving flows]
$ $
\label{asm_flow}
\begin{enumerate}
    \item[\ref{asm_flow}.1] Sending flows: For every $e\in\mathcal{E}$, $f_e(x_e)$ is Lipschitz continuous and non-decreasing in $x_e$. Furthermore, $f_e(x_e)=0$ if $x_e=0$ and $\sup_{x_e} f_e(x_e) < \infty$.
    \item[\ref{asm_flow}.2] Receiving flows: For every $e\in\mathcal{E}$, $ r_e(x_e)$ is Lipschitz continuous and non-increasing in $x_e$. Furthermore, for link $e$ with finite storage space, we assume $r_e(x_e) = 0$ if $x_e=x_e^{\max}$; for link $e$ with infinite storage space, we assume a constant receiving flow $r_e$ independent of $x_e$.
\end{enumerate}
\end{asm}

Note that receiving flows are not considered in some flow networks. In that case we just let link $e$ have infinite storage space with $r_e=\infty$. 

We define \emph{link capacity} $Q_e$ and \emph{critical density} $x_e^c$ as
\begin{subequations}
\begin{align}
    Q_e &:= \sup_{x_e\in\mathcal X_e}\min\{f_e(x_e),r_e(x_e)\}, \label{eq_def_Fk} \\
    x_e^c &:= \inf\{x_e\in\mathcal{X}_e|f_e(x_e) = Q_e\}. \label{eq_def_xkc}
\end{align}
\end{subequations}
Practically, the capacity $Q_e$ indicates an upper bound of sustainable outflow from link $e$, and the critical density $x_e^c$ denotes an threshold where the capacity flow $Q_e$ can be maintained with a relatively high speed $Q_e/x_e^c$. Link $e$ is considered as ``congested'' when its density $x_e$ exceeds the critical value $x_e^c$.


Here are two examples of sending and receiving flows. In road network, as per the cell transmission model (CTM \cite{daganzo95}), the sending and receiving flows are given by:
\begin{subequations}
    \begin{align}
        f_e(x_e)&=\min\{v_fx_e,Q_e\}, \label{eq_asm_exm_ctm_f} \\
        r_e(x_e)&= \min\{Q_e,w_c(x_e^{\max}-x_e)\}, \label{eq_asm_exm_ctm_r}
    \end{align}
\end{subequations}
where $x_e$ represents vehicle density of road section $e$, and traffic parameters $v_f$, $Q_e$, $w_c$ and $x_e^{\max}$ are typically assumed to 
satisfy $Q_e/x_e^{\max}\leq v_fw_c/(v_f+w_c)$ \cite{jin2018analysis}. It follows $x_e^c=Q_e/v_f$. In data networks, the sending flow can be approximated as a fluid model \cite{como13i} with
    \begin{equation}
        f_e(x_e)=Q_e(1-e^{-\rho_e x_e}), \label{eq_asm_exm_data_f}
    \end{equation}
where $x_e$ stands for queue length on channel $e$, $\rho_e$ is a positive constant, and $Q_e$ is channel capacity. 
The receiving flow is not explicitly modeled in data networks and thus we assume $r_e=\infty$, which yields $x_e^c=\infty$.

\begin{figure}[h]
    \centering
    \begin{subfigure}{0.3\linewidth}
        \centering
        \includegraphics[width=\linewidth]{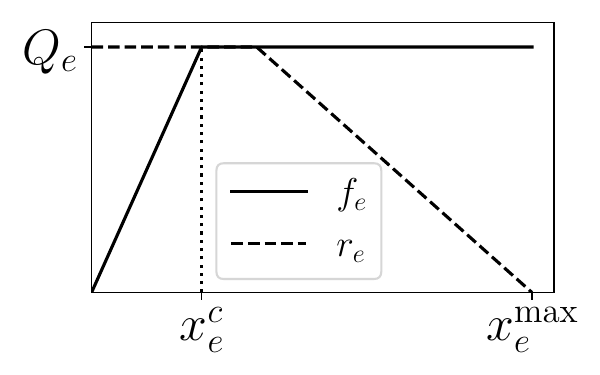}
        \caption{Road networks.}
    \end{subfigure}
    \begin{subfigure}{0.3\linewidth}
        \centering
        \includegraphics[width=\linewidth]{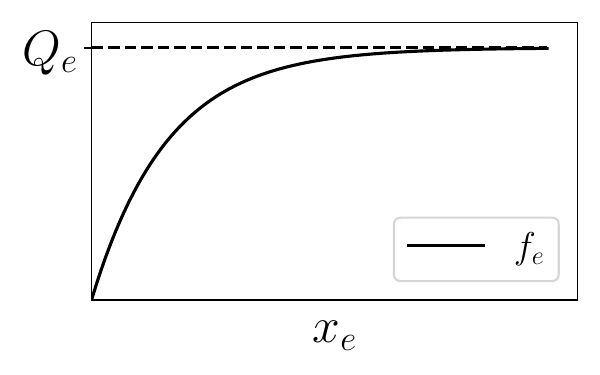}
        \caption{Data networks.}
    \end{subfigure}
    \caption{Examples of sending/receiving flows.}
    \label{fig_ex_flow}
\end{figure}

\subsubsection{Control laws}
We consider a control law $\mu:\mathcal X\to\mathbb R_{\ge0}^{\mathcal{P}}$ to be a concatenation of controllers $\mu_{ej}:\mathcal{X}\to\mathbb{R}_{\geq0}$, $(e,j)\in\mathcal{P}$, that regulate flows from link $e$ to link $j$. In this paper we focus on locally responsive control laws defined below.
\begin{dfn}[Locally responsive control laws]
A control law $\mu$ for a network $\mathcal{G}=(\mathcal{V},\mathcal{E})$ is \emph{locally responsive} if $\mu_{ej}(x)$, for any $(e,j)\in\mathcal{P}$, can depend on the states of the links in $\mathcal{E}_{\tau_e}^-\cup\mathcal{E}_{\tau_e}^+$ only, where $\mathcal{E}_{\tau_e}^-\cup\mathcal{E}_{\tau_e}^+$ denotes the set of incoming and outgoing links of node $\tau_e$.
\end{dfn}

In general, \emph{actual inter-link flows} $q^{\mu}:\mathcal{X}\to\mathbb{R}_{\geq0}^{\mathcal{P}}$ may not equal the control inputs $\mu(x)$ if the control law violates sending or receiving flows, e.g. under open-loop control. We define it as follows:
\begin{equation}
    q_{ej}^{\mu}(x) := \min\{\mu_{ej}(x), f_{ej}^{\mu}(x),
    r_{ej}^{\mu}(x)\}, 
    \label{eq_q}
\end{equation}
where
\begin{subequations}
    \begin{align}
        f^{\mu}_{ej}(x) :=& \frac{\mu_{ej}(x)}{\sum_{j'\in\mathcal{E}_e^+}\mu_{ej'}(x)}f_{e}(x_e), \label{eq_f_ej} \\
        r^{\mu}_{ej}(x) :=& \frac{\mu_{ej}(x)}{\sum_{e'\in\mathcal{E}_j^-}\mu_{e'j}(x)}r_j(x_j). \label{eq_r_ej}
    \end{align}
\end{subequations}
This modelling approach enables us to study inappropriate control instructions induced by cyber-physical disruption introduced later. It is necessary since controllers may not have accurate knowledge of sending/receiving flows due to disruptions. The actual controlled flow $q_{ej}^{\mu}(x)$ represents how the network responds to the control input $\mu$. 
The sending and receiving flows are allocated proportionally to $\mu_{ej}(x)$, which is a typical modeling approach; for more discussion on this and alternative models, see \cite{kurzhanskiy10}. Obviously, these techniques, with minor modifications, can be applied to flow allocations in multi-commodity dynamical flow networks \cite{nilsson2014resilience}. Note that the superscript ``$\mu$'' in \eqref{eq_q} indicates the reliance of inter-links flows on the control law. We typically omit it unless we want to emphasize the dependence on a particular control law.

\begin{asm}[Controlled flows]
\label{asm_policy}
$ $
\begin{enumerate}
    \item[\ref{asm_policy}.1] Continuity: The controlled flow $q(x)$ is Lipschitz continuous in $x_e$ for any $e\in\mathcal{E}$.
    \item[\ref{asm_policy}.2] Monotonicity: For any two different links $e,e'\in\mathcal{E}$, 
    \begin{enumerate}[(i)]
        \item $\sum_{i\in\mathcal{E}_e^-}q_{ie}(x)$ is non-decreasing in $x_{e'}$;
        \item $\sum_{j\in\mathcal{E}_e^+}q_{ej}(x)$ is non-increasing in $x_{e'}$.
    \end{enumerate}
\end{enumerate}
\end{asm}

Assumption \ref{asm_policy}.1 indicates that our approach is applicable to non-smooth control laws. Assumption \ref{asm_policy}.2 implies that the considered dynamic flow network is cooperative, which is a special class of monotone systems \cite{hirsch85}. The following elaborates Assumption 2.2 (i) to reveal the cooperativity. If link $e'$ locates upstream of link $e$ and $x_{e'}$ increases, Assumption 2.2 (i) implies that the link set $\mathcal{E}_{e}^-$ tends to send more flows to link $e$ in order to avoid density accumulation in the upstream. If $e'$ does not appear upstream of link $e$ and $e'\neq e$, the increase of $x_{e'}$ implies that somewhere (at least not link $e$) becomes congested and Assumption 2.2 (i) indicates that the link set $\mathcal{E}_{e}^-$ tends to send more flows to link $e$, which benefits alleviating congestion somewhere.  A similar explanation applies to Assumption 2.2 (ii). Thus Assumption 2.2 indicates how the links reduce congestion cooperatively.

While dynamic flow networks are, in general, not monotone \cite{coogan15}, the rationale behind Assumption \ref{asm_policy}.2 lies in that the monotonicity applies to a broad class of flow networks \cite{como2017resilient} and provides a tractable structure that benefits analysis and computation. The assumption of monotonicity can be relaxed in our stability analysis, with a rise of computational costs; see Propositions~\ref{prp_stable} and \ref{prp_stable_general}.

%
%

Below are the examples regarding routing and merging strategies satisfying Assumption \ref{asm_policy}. Consider routing control for a diverging junction with an upstream link $e$. The classical logit routing policy \cite{como13i} gives
\begin{equation}
    q_{ej}(x)=\min\Big\{\frac{e^{-\theta_jx_j}}{\sum_{j'\in\mathcal{E}^+_e}e^{-\theta_{j'} x_{j'}}}f_e(x_e), r_j(x_j)\Big\},
\end{equation}
where $\theta_j$ quantifies the sensitivity of route choice to $x_j$. Now consider a merging junction with a minor upstream link $e_{i_1}$, a major upstream link $e_{i_2}$ and a downstream link $e$. A typical ramp metering policy, namely the occupancy strategy \cite{gomes2003study}, yields
\begin{subequations}
    \begin{align}
        q_{e_{i_1}e}(x) &= \min\{u-\kappa x_{e_{i_2}}, f_{e_{i_1}}(x_{e_{i_1}}), r_e(x_e)\}, \label{eq_asm_exm_rm_1} \\
        q_{e_{i_2}e}(x) &= \min\{r_e(x_e)-q_{e_{i_1}e}(x), f_{e_{i_2}}(x_{e_{i_2}})\}, \label{eq_asm_exm_rm_2}
    \end{align}
\end{subequations}
where $u,\kappa\geq0$ are control parameters.

\subsubsection{Disruption modes}
We consider a set $\mathcal S$ of \emph{modes} that capture cyber and/or physical disruptions. 
With a slight abuse of notation, we use $f_e:\mathcal{S}\times\mathcal{X}_e\to\mathbb{R}_{\ge0}$,  $r_e:\mathcal{S}\times\mathcal{X}_e\to\mathbb{R}_{\ge0}$, $\mu:\mathcal S\times\mathcal X\to\mathbb R_{\ge0}^{\mathcal{P}}$ and $q:\mathcal S\times\mathcal X\to\mathbb R_{\ge0}^{\mathcal{P}}$ to denote the sending flow of link $e$, the receiving flow of link $e$, the control law and the actual inter-link flows influenced by disruptions, respectively. 

\begin{asm}[Cyber-physical disruptions]
\label{asm_fault}
$ $
\begin{enumerate}
    \item[\ref{asm_fault}.1] Nominal mode: There exists a nominal mode, denoted by $s_0\in\mathcal{S}$, under which the network is free from cyber-physical disruptions and thus stable.
    \item[\ref{asm_fault}.2] Disrupted sending/receiving flows: For every $s\in\mathcal{S}$ and $e\in\mathcal{E}$, $f_e(s,x_e)$ and $r_e(s,x_e)$ satisfy Assumption~\ref{asm_flow}. Furthermore,  $f_e(s,x_e)\leq f_e(s_0,x_e)$ and $r_e(s,x_e)\leq r_e(s_0,x_e)$.
    \item[\ref{asm_fault}.3] Disrupted controlled flows: For every $s\in\mathcal{S}$ and $e\in\mathcal{E}$,
    $q(s,x)$ satisfies Assumption \ref{asm_policy}.
\end{enumerate}
\end{asm}

Assumption~\ref{asm_fault}.1 ensures one stable mode. In this paper, our primal interest is to study whether the network is destabilized given the switched modes. Assumptions~\ref{asm_fault}.2 and \ref{asm_fault}.3 imply that disruptions will not fundamentally change the flow dynamics. They hold for typical reliability failures\footnote{Security failures, which are beyond the scope of this paper, however, may violate this assumption.}; see the examples later. Then mode-specific link capacities and critical densities are given by
\begin{subequations}
\begin{align}
    Q_{se} :=& \sup_{x_e\in\mathcal X_e}\min\{f_e(s,x_e),r_e(s,x_e)\}, \\
    x_{se}^c :=& \inf \{x_e\in\mathcal{X}_e|f_e(s,x_e) = Q_{se}\}. \label{eq_mode_crti}
\end{align}
\end{subequations}
Note that Assumption~\ref{asm_fault}.2 also indicates $Q_{se}\leq Q_{s_0e}$ for any $s\in\mathcal{S}$ and $e\in\mathcal{E}$. 
We denote by $x_e^{c*}$ the maximum critical density over disruption modes:
\begin{equation}
    x_e^{c*} := \max_{s\in\mathcal{S}} x_{se}^c,~e\in\mathcal{E}.
    \label{eq_x_e_c}
\end{equation}


The following are examples of physical and cyber disruptions. Consider logit routing for a diverging junction with an upstream link $e$, and suppose that the outflow from link $e$ may decrease by a certain ratio. Thus we consider $\mathcal{S}=\{s_0, s_1\}$ and $f_e(s_1,x_e)=\chi f_e(s_0,x_e)$, where $\chi\in[0, 1)$. The actual inter-link flows are given by
\begin{equation}
    q_{ej}(s,x)=\min\Big\{\frac{e^{-\theta_jx_j}}{\sum_{j'\in\mathcal{E}^+_e}e^{-\theta_{j'} x_{j'}}}f_e(s,x_e), r_j(x_j)\Big\}.
\end{equation}
Now suppose that the observation of link $j^*\in\mathcal{E}_e^+$, denoted by $T_{j^*}(s,x)$, may be biased. Without loss of generality, we let $T_{j^*}(s_1,x):=\kappa x_{j^*}$ with some $\kappa\geq0$ and $T_{j'}(s,x)= x_{j'}$ for any $(s,j')\neq(s_1,j^*)$. The controlled flows are given by
\begin{equation}
    q_{ej}(s,x)=\min\Big\{\frac{e^{-\theta_jT_j(s,x)}}{\sum_{j'\in\mathcal{E}^+_e}e^{-\theta_{j'} T_{j'}(s,x) }}f_e(x_e), r_j(x_j)\Big\}.
\end{equation}

\subsection{PDMP model}
\label{sub_pdmp}

Now  we can define the piecewise-deterministic dynamics of the controlled process $\{(S(t),X(t));t\ge0\}$. The discrete-state process $\{S(t);t\ge0\}$ of the mode is a homogeneous finite-state Markov process that is independent of the continuous-state process $\{X(t);t\ge0\}$ of the traffic densities. The state space of the discrete process is $\mathcal S:=\{s_0,s_1,\cdots,s_M\}$. The \emph{transition rate} from mode $s$ to mode $s'$ is $\lambda_{ss'}$. Without loss of generality, we assume that $\lambda_{ss}=0$ for all $s\in\mathcal S$. The discrete-state process evolves as follows:
$$
\Pr\{S(t+\delta)=s'|S(t)=s\}=\lambda_{ss'}\delta+\mathrm o(\delta),~\forall s, s' \in\mathcal S,
$$
where $\delta$ denotes an infinitesimal increment. We assume that the discrete-state process is ergodic, which means that every disruption will be resolved in finite time almost surely. It follows that $\{S(t);t\geq0\}$ admits a unique steady-state probability vector $p:=[p_{s_0},p_{s_1},\cdots,p_{s_M}]^{\mathrm{T}}\in\mathbb{R}_{\geq0}^{\mathcal{S}}$ satisfying 
\begin{equation}
    \Lambda^{\mathrm{T}}p = \bm{0}, \label{eq_p}
\end{equation}
where $\bm{0}$ denotes the zero vector and
$\Lambda$ is the transition matrix given by
\begin{equation}
\Lambda := \begin{bmatrix}
    -\sum\limits_{s'\in\mathcal{S}}\lambda_{s_0s'} & \lambda_{s_0s_1} & \cdots & \lambda_{s_0s_M} \\
    \lambda_{s_1s_0} & -\sum\limits_{s'\in\mathcal{S}}\lambda_{s_1s'} & \cdots & \lambda_{s_2s_M} \\
    \vdots & \vdots & \ddots & \vdots \\
    \lambda_{s_Ms_0} & \lambda_{s_Ms_1} & \cdots & -\sum\limits_{s'\in\mathcal{S}}\lambda_{s_Ms'}
    \end{bmatrix}.
\end{equation}

The continuous-state process $\{X(t);t\ge0\}$ is defined as follows. We let $G_e(s,x):=(\mathrm{d}/\mathrm{d}t)X_e(t)$. 
The conservation law associated with flows implies
\begin{subequations}
\begin{align}
&G_e(S(t),X(t))=\quad\quad\quad\quad\quad\quad\quad\quad\quad\quad\quad\quad\quad\quad\quad\quad \nonumber
\end{align}
\begin{numcases}{}
\alpha-\sum_{j\in\mathcal{E}^+_{e}}q_{ej}(S(t),X(t)), & if $\tau_e=v_o$  \label{eq_Ga} \\
\sum_{i\in\mathcal{E}^-_e}q_{ie}(S(t),X(t)) & \nonumber \\  
\quad\quad -\sum_{j\in\mathcal{E}^+_e}q_{ej}(S(t),X(t)), & if $\tau_e\notin\{v_o,v_d\}$  \\
\sum_{i\in\mathcal{E}^-_e}q_{ie}(S(t),X(t)) & \nonumber \\
\quad\quad - f_e(S(t), X(t)), & if $\tau_e=v_d$. \label{eq_Gc}
\end{numcases} 
\end{subequations}
Note 
$G_e$ is bounded. So is $G(s,x):=(\mathrm{d}/\mathrm{d}t)X(t)$.

The joint evolution of $S(t)$ and $X(t)$ is a PDMP and can be described compactly using an infinitesimal generator \cite{meyn93}
\begin{align}
    &\mathscr L V(s,x) =\nonumber \\
    & G(s,x)^{\mathrm{T}}\nabla_{x}V(s,x) + \sum\limits_{s'\in\mathcal S}\lambda_{ss'}(V(s',x)-V(s,x))
    \label{eq_infinitesimal}
\end{align}
for any differentiable function $V:\mathcal S\times\mathcal X\to\mathbb R_{\geq0}$, where $\nabla_{x}V(s,x)$ is the gradient of $V$ with respect to $x$. The bounded dynamics $G(s,x)$ indicates that $\{(S(t),X(t));t\geq0\}$ is a \emph{non-explosive} Markov process, which is a prerequisite for discussing stability \cite{meyn93}.

\subsection{Stability and resiliency}
\label{sub_add}

In this subsection, we define network stability and resiliency, which are the focus of the subsequent analysis.

\subsubsection{Stability and invariant set}

Below we present several concepts regarding stability.

\begin{dfn}[Stability \& Instability]
The network is \emph{stable} if there exists a scalar $Z<\infty$ such that for any initial condition $(s,x)\in\mathcal S\times\mathcal X$
\begin{equation}\label{eq_bounded}
  \limsup_{t\to\infty}\frac1t\int_{\tau=0}^t\mathbb E[|X(\tau)||S(0)=s,X(0)=x]\mathrm{d}\tau\le Z,
\end{equation}
where $|X(t)|$ denotes 1-norm of $X(t)$, namely $|X(t)|=\sum_{e\in\mathcal{E}}|X_e(t)|$. The network is \emph{unstable} if there does not exist $Z<\infty$ such that for any initial condition $(s,x)\in\mathcal{S}\times\mathcal{X}$ such that the inequality \eqref{eq_bounded} holds.
\end{dfn}

The notion of stability follows a classical definition \cite{dai95ii}, which is closely related to ``first-moment stable'' \cite{shi2015survey}.
Practically, if the time-average traffic densities in all links are bounded, the network is stable; otherwise, it is unstable.

Verifying the stability requires to check \eqref{eq_bounded} for all initial conditions $(s,x)\in\mathcal{S}\times\mathcal{X}$.  This can be simplified by considering an \emph{invariant set} $\mathcal X^{\mu}\subseteq\mathcal{X}$ \cite{benaim15}, which is defined as follows:
\begin{dfn}[Invariant set] For the PDMP $\{(S(t),X(t));t\geq0\}$, a subset $\mathcal{X}^{\mu}\subseteq\mathcal{X}$ is an invariant set if it is (i) globally attracting and (ii) positively invariant:
\begin{align*}
    (\mathrm{i})&~\forall (S(0),X(0))\in\mathcal{S}\times\mathcal{X}, \\
    &\quad \lim_{t\to\infty}\mathrm{Pr}\{X(t)\in\mathcal{X}^\mu|(S(0),X(0))\} = 1; \\
    (\mathrm{ii})&~\forall (S(0),X(0))\in\mathcal{S}\times\mathcal{X}^\mu,\forall t\geq0, X(t)\in \mathcal{X}^\mu.
\end{align*}
\end{dfn}
The definition above indicates that i) for any initial condition $X(0)\in\mathcal{X}$, the state $X(t)$ enters $\mathcal{X}^\mu$ almost surely, and that ii) given any initial condition $X(0)\in\mathcal{X}^\mu$, the state $X(t)$ never leaves $\mathcal{X}^\mu$. Note that one trivial candidate set is $\mathcal{X}$ itself,  but a tight invariant set depends on the control law $\mu$ and disruption modes $\mathcal{S}$. It is assumed that $\mathcal{X}^\mu$ is unbounded; otherwise, the network is naturally stable. Besides, we define
\begin{equation*}
    \underline x_e^{\mu}:=\inf_{x\in\mathcal X^\mu}x_e,~\bar x_e^{\mu}:=\sup_{x\in\mathcal X^\mu}x_e, ~e\in\mathcal E.
\end{equation*}
If $x_e$ does not have an invariant upper bound, we let $\bar{x}_e^\mu=\infty$. Clearly, we have $\mathcal{X}^\mu\subseteq\prod_{e\in\mathcal{E}}[\underline{x}_e^\mu, \bar{x}_e^\mu]$.

The following sets of links are induced by the invariant set $\mathcal{X}^{\mu}$. Clearly they also depend on the control law $\mu$, but we omit the superscript ``$\mu$'' for notational convenience without causing confusion.
We denote by $\mathcal{E}_{\mathrm{inf}}:=\{e\in\mathcal{E}| \bar{x}_e^\mu = \infty\}$ the set of density-unbounded links. For any $e\in\mathcal{E}_{\mathrm{inf}}$, let $\mathcal{A}^{-}_{e,\mathrm{inf}}:=\mathcal{A}_e^-\cap\mathcal{E}_{\mathrm{inf}}$ (resp. $\mathcal{A}^{+}_{e,\mathrm{inf}}:=\mathcal{A}_e^+\cap\mathcal{E}_{\mathrm{inf}}$) denote its upstream (resp. downstream) density-unbounded links. We also define the set of density-bounded links that are upstream (resp. downstream) of link $e\in\mathcal{E}_{\mathrm{inf}}$, denoted by $\mathcal{A}_{e,\mathrm{fin}}^{-} := \mathcal{A}_e^{-}\setminus\mathcal{A}_{e,\mathrm{inf}}^{-}$ (resp. $\mathcal{A}_{e,\mathrm{fin}}^{+} := \mathcal{A}_e^{+}\setminus\mathcal{A}_{e,\mathrm{inf}}^{+}$). Particularly, we let
\begin{align}
 \mathcal{B}_{e}^+ := \{n\in\mathcal A_e^+| \forall  \ell & \in\{n\} \cup(\mathcal{A}_{e}^+\cap\mathcal{A}_{n}^-), \nonumber \\ 
~(i)&~ \ell\in\mathcal{A}_{e,\mathrm{fin}}^+, \nonumber \\
~(ii)&~ \exists (s,x)\in\mathcal{S}\times\mathcal{X}^{\mu},\exists i\in\mathcal{E}_{\ell}^-, \nonumber \\
&\quad 
\min\{\mu_{i\ell}(s,x), f^\mu_{i\ell}(s,x)\} > r^\mu_{i\ell}(s,x)\} \label{eq_B+}
\end{align}
denote the set of links downstream of link $e\in\mathcal{E}_{\mathrm{inf}}$ such that (i) link $n\in\mathcal{B}_e^+$ is density-bounded, (ii) link $n\in\mathcal{B}_e^+$ can block the flows from its upstream link $i$, and (iii) any link $\ell$ between $e$ and $n$, namely $\ell\in\mathcal{A}_e^+\cap\mathcal{A}_n^-$ satisfies (i) and (ii), where $f^\mu_{i\ell}(s,x)$ (resp. $r^\mu_{i\ell}(s,x)$), similar to \eqref{eq_f_ej} (resp. \eqref{eq_r_ej}), denotes allocated sending (resp. receiving) flow between links $i$ and $\ell$. Intuitively, $\mathcal{B}_{e}^{+}$ is the set of connected bottlenecks, between link $e$ and its downstream density-unbounded links, that induce congestion spillback and block the discharging flow of link $e$.

We use the classical Wheatstone bridge \cite{ziliaskopoulos00} to illustrate the definitions above; see Fig.~\ref{fig_network}. Suppose that only links $e_o$ and $e_5$ have infinite storage and that the flows out of links $e_2$ and $e_5$ are decreased randomly by physical disruptions. Then the congestion builds up at links $e_2$ and $e_5$. Due to the infinite storage, the traffic density at link $e_5$ may blow up. Meanwhile, link $e_o$ could be blocked as well once the jam at link $e_2$ spills back to link $e_o$ via link $e_1$. It indicates $\mathcal{E}_{\mathrm{inf}}=\{e_o,e_5\}$ with $\mathcal{B}_{e_o}^+=\{e_1,e_2\}$ and $\mathcal{B}_{e_5}^+=\varnothing$.
\begin{figure}[htbp]
    \centering
    \begin{subfigure}{0.45\linewidth}
    \resizebox{\linewidth}{!}{
    \begin{tikzpicture}
        \Vertex[x=-0.9,style={color=white}]{A}
        \Vertex[label=$v_o$,x=1,fontscale=1]{B}
        \Vertex[label=$v_1$,x=2.5,y=1,fontscale=1]{C}
        \Vertex[label=$v_2$,x=2.5,y=-1,fontscale=1]{D}
        \Vertex[label=$v_d$,x=4,fontscale=1]{E}
        \Edge[Direct,label=$e_o$,fontscale=1,style=dotted](A)(B)
        \Edge[Direct,label=$e_1$,fontscale=1](B)(C)
        \Edge[Direct,label=$e_2$,fontscale=1](C)(D)
        \Edge[Direct,label=$e_3$,fontscale=1](C)(E)
        \Edge[Direct,label=$e_4$,fontscale=1](B)(D)
        \Edge[Direct,label=$e_5$,fontscale=1,style=dotted](D)(E)
        
        \draw[dashed] (1.4,-1.4) -- (1.4,1.7) 
        -- (4.9,1.7) -- cycle;
        \Vertex[x=3.6,y=1.7,size=1,style={white},label=$\mathcal{A}_{e_o,\mathrm{fin}}^+$,fontscale=1]{A_plus_fin}
        
        \draw[dashed] (1.8,0.9) to[bend right=45] (2.8,0.4);
        \Vertex[x=1.8,y=1.2,size=0.2,style={white},label=$\mathcal{B}_{e_o}^+$,fontscale=1]{B_plus}
        
        \draw[dashed] (3.2,0) -- (3.7, -0.8);
        \Vertex[x=4,y=-1.1,size=0.2,style={white},label=$\mathcal{A}_{e_o,\mathrm{inf}}^+$,fontscale=1]{A_plus_inf}
        
    \end{tikzpicture}
    }
    \caption{ $\mathcal{A}_{e_o,\mathrm{inf}}^+$, $\mathcal{A}_{e_o,\mathrm{fin}}^+$ and $\mathcal{B}_{e_o}^+$.}
    \end{subfigure}
    \begin{subfigure}{0.42\linewidth}
    \resizebox{\linewidth}{!}{
    \begin{tikzpicture}
        \Vertex[x=-0.9,style={color=white}]{A}
        \Vertex[label=$v_o$,x=1,fontscale=1]{B}
        \Vertex[label=$v_1$,x=2.5,y=1,fontscale=1]{C}
        \Vertex[label=$v_2$,x=2.5,y=-1,fontscale=1]{D}
        \Vertex[label=$v_d$,x=4,fontscale=1]{E}
        \Edge[Direct,label=$e_o$,fontscale=1,style=dotted](A)(B)
        \Edge[Direct,label=$e_1$,fontscale=1](B)(C)
        \Edge[Direct,label=$e_2$,fontscale=1](C)(D)
        \Edge[Direct,label=$e_3$,fontscale=1](C)(E)
        \Edge[Direct,label=$e_4$,fontscale=1](B)(D)
        \Edge[Direct,label=$e_5$,fontscale=1,style=dotted](D)(E)
        
        \draw[dashed] (1.4,-1.4) -- (1.4,1.6) 
        -- (2.9,0.1) -- cycle;
        \Vertex[x=2.2,y=1.7,size=0.1,style={white},label=$\mathcal{A}_{e_5,\mathrm{fin}}^-$,fontscale=1]{A_minus_fin}
        
        \draw[dashed] (-0.43,-0.5) -- (-0.43,0.5);
        \Vertex[x=-0.1,y=0.7,size=0.1,style={white},label=$\mathcal{A}_{e_5,\mathrm{inf}}^-$,fontscale=1]{A_minus_inf}
        
    \end{tikzpicture}
    }
    \caption{ $\mathcal{A}_{e_5,\mathrm{inf}}^-$ and $\mathcal{A}_{e_5,\mathrm{fin}}^-$.}
    \end{subfigure}
    
    \caption{Wheatstone bridge with $\mathcal{E}_{\mathrm{inf}}=\{e_o,e_5\}$ denoted by dotted lines: $\mathcal{B}_{e_o}^+=\{e_1,e_2\}$, $\mathcal{A}_{e_o,\mathrm{inf}}^+=\{e_5\}$, $\mathcal{A}_{e_o,\mathrm{fin}}^+=\{e_1,e_2,e_3,e_4\}$, $\mathcal{A}_{e_5,\mathrm{inf}}^-=\{e_o\}$ and $\mathcal{A}_{e_5,\mathrm{fin}}^-=\{e_1,e_2,e_4\}$.}
    \label{fig_network}
\end{figure}


\subsubsection{Resiliency, throughput and min-cut capacities}
The \emph{throughput} $\alpha^\mu$ of a network with control $\mu$ is defined as the maximal demand that the network can accept while maintaining stability, i.e. traffic densities in all links being bounded on average:
\begin{align}
    \alpha^{\mu}:=\sup\alpha\quad \text{s.t.}\mbox{ network is stable.}
\end{align}
%

For a network $\mathcal{G}=(\mathcal{E},\mathcal{V})$ with link capacities $\mathcal{Q}:=\{Q_e|e\in\mathcal{E}\}$, we denote by $C(\mathcal{Q};\mathcal{G})$ its min-cut capacity (MCC). The min-cut max-flow theorem states that the nominal throughput equals the MCC \cite{dantzig1955max}, which is obtained by solving the maximum flow problem:
\begin{subequations}
\begin{align}
  \mbox{($\mathrm{P}_1$)}~ &\max_{\alpha\geq0,u\in\mathbb R_{\ge0}^{\mathcal{P}}}~\alpha \nonumber\\
    \mbox{s.t.}& ~ \alpha = \sum_{j\in\mathcal{E}_e^+}u_{ej},~\tau_e=v_o,   \label{eq_p1_1} \\
    &~ \sum_{i\in\mathcal{E}_e^-}u_{ie} = \sum_{j\in\mathcal{E}_e^+}u_{ej}, ~\tau_e\in\mathcal{V}\setminus\{v_o,v_d\},  \label{eq_p1_2} \\
    &~\sum_{i\in\mathcal{E}^-_e}u_{ie} \le Q_e,~\tau_e\in\mathcal{V}\setminus\{v_o\},  \label{eq_p1_3} \\
    &~\sum_{j\in\mathcal{E}^+_e}u_{ej} \le Q_e,~\tau_e\in\mathcal{V}\setminus\{v_d\}, \label{eq_p1_4}
\end{align}
\end{subequations}
where the optimal $\alpha^*$ is equal to $C(\mathcal{Q};\mathcal G)$. Then we define  resiliency score as
\begin{equation}
\label{eq_resiliencyscore}
\eta^\mu := \alpha^\mu / C(\mathcal{Q}_{s_0};\mathcal{G}),
\end{equation}
where $C(\mathcal{Q}_{s_0};\mathcal{G})$ denotes the MCC in the nominal mode $s_0$. Assumption~\ref{asm_fault}.2 indicates $\eta^\mu\in[0, 1]$. If $\eta^\mu=1$, we say that the control law $\mu$ is strongly resilient against disruptions.  The motivation for the resiliency score lies in that min-cut capacity is an important measurement of network capability. Meanwhile, for monotone dynamical flow networks, it is easily attainable with an open-loop control in the nominal case; see Appendix~\ref{app_olnominal}. Thus the ratio can capture throughput losses due to disruptions.


\section{Resiliency analysis}
\label{sec_analysis}

In this section, we analyze a network's stability under a given control law $\mu$. The main results allow quantification of the resiliency score $\eta^\mu$ given by \eqref{eq_resiliencyscore}.


To state the results, we define
\begin{subequations}
\begin{align}
I_{e}(s, x) :=& \sum_{m\in\mathcal A_e^{-}} G_{m}(s,x) + \sum_{i\in\mathcal{E}_e^-}q_{ie}(s,x), \label{eq_in} \\
O_e(s, x) :=& \sum_{j\in\mathcal{E}_e^+}q_{ej}(s,x)-\sum_{\ell\in\mathcal B_{e}^{+}}\rho_{\ell}(x_\ell)G_\ell(s,x),
\label{eq_out}
\end{align}
\end{subequations}
where $\rho_\ell:\mathbb R_{\ge0}\to\mathbb R_{\ge0}$ is a weight function:
\begin{align}
\rho_{\ell}(x_{\ell})=\begin{cases}
0 & x_{\ell} < \underline x_{\ell}^\mu, \\
\frac{x_{\ell}-\underline x_{\ell}^\mu}{\bar{x}_\ell^\mu - \underline x_{\ell}^\mu} & \underline x_\ell^\mu \le x_\ell<\bar{x}_\ell^\mu, \\
1 &  x_\ell \ge \bar x_\ell^\mu.
\end{cases}
\label{eq_rho}
\end{align}
The weight function indicates that $G_\ell$, the dynamics of link $\ell$, has more impacts on \eqref{eq_out} as the traffic density $x_\ell$ increases. We interpret \eqref{eq_in} as the inflow from the origin $v_o$ to link $e$, by considering the flow conservation but not the dynamics on the
intermediate links. To see that, we first notice by \eqref{eq_Ga}-\eqref{eq_Gc} that $G_m(s,x)$ is the net flow only considering neighbor links of link $m$, and then recognize $\sum_{m\in\mathcal{A}_e^-}G_m(s,x)$ as the net flow of the upstream network whose inflow is the demand $\alpha$ and whose outflows comprise of i) those passing link $e$ and ii) those never passing link $e$. Since the former outflows are canceled by  $\sum_{i\in\mathcal{E}_e^-}q_{ie}(s,x)$ in \eqref{eq_in}, it is easy to understand that \eqref{eq_in} denotes the demand $\alpha$ minus the flows never passing link $e$. In a similar way, we interpret \eqref{eq_out} as a weighted outflow of link $e$ including those traversing the downstream bottlenecks and those not. Aware of the physical meanings of \eqref{eq_in}-\eqref{eq_out}, we let
\begin{equation}
    N_{e}(s,x) := I_e(s,x) - O_e(s,x) \label{eq_net}
\end{equation}
denote the net flow of link $e$ considered in a larger scope, from the location of demand generation to the sites of bottleneck dissipation. It captures necessary networkwide dynamics and thus helps investigate the network stability conditions.

We use the Wheatstone bridge in Fig.~\ref{fig_network} to enhance the understandings of \eqref{eq_in}-\eqref{eq_out}. First, we have $I_{e_5}(s,x)=\alpha-q_{e_1e_3}(s,x)$, where the flow $q_{e_1e_3}$ never enters link $e_5$. If we plug $\rho_\ell\equiv1$ into \eqref{eq_out}, we obtain
$O_{e_o}(s,x)= q_{e_oe_4}(s,x) + q_{e_1e_3}(s,x) + q_{e_2e_5}(s,x) $, where $q_{e_2e_5}$ denotes the outflow of link $e_o$ through the bottleneck and $q_{e_oe_4}+q_{e_1e_3}$ represents the flows never entering the bottleneck. Although the physical meaning is clear given $\rho_\ell\equiv1$, it leads to trivial stability analysis. This is because congestion spillback is oversimplified. For example, the flows $q_{e_oe_1}$ and $q_{e_1e_2}$ are omitted in $O_{e_o}(s,x)$. This will happen if piecewise quadratic Lyapunov functions with $\rho_\ell\equiv1$ are used.

%
%

\subsection{Main results}
This subsection presents two main results of stability analysis. The first stability condition emphasizes a physical intuition that any dynamic flow network is stable if the long-term net flow is negative for each density-unbounded link; the second is stronger but more abstract.

The first main result is as follows:
\begin{thm}
\label{thm_stable}
Consider an acyclic network satisfying Assumptions~\ref{asm_fault}.1-\ref{asm_fault}.3. Suppose that the network admits a demand $\alpha$ and an invariant set $\mathcal X^\mu\subseteq\prod_{e\in\mathcal{E}}[\underline{x}_e^\mu, \bar{x}_e^\mu]$ under a control law $\mu:\mathcal S\times\mathcal X\to\mathbb R_{\ge0}^{\mathcal{P}}$. Then, the network is stable if
    \begin{equation}
        \sum_{s\in\mathcal S}p_s\max_{x\in\mathcal{D}_e^\mu} 
        N_e(s,x)<0,~\forall e\in \mathcal{E}_{\mathrm{inf}}, \label{eq_psG}
    \end{equation}
    where $\{p_{s}|s\in\mathcal{S}\}$ is the steady-state probability distribution of disruption modes and $\mathcal{D}_e^\mu$ is given by
    \begin{align}
        \mathcal{D}_e^\mu:=\{x|&x_e=x_e^{c*},~x_m=\underline{x}_m^\mu,~\forall m \in \mathcal{A}_{e}^{-}, \nonumber \\
        &x_n=\bar{x}_n^\mu,~\forall n\in\mathcal{E}\setminus(\mathcal{A}^-_{e}\cup\{e\}\cup\mathcal{B}_e^+), \nonumber \\
        & x_\ell\in[\underline{x}_\ell^\mu,\bar{x}_\ell^\mu], ~\forall \ell\in\mathcal{B}_e^+\}. \label{eq_compact_D}
\end{align}
\end{thm}

Theorem~\ref{thm_stable} essentially states that the network is stable if the expectation of the maximum $N_e(s,x)$ 
over $\mathcal{D}_e^\mu$ 
is negative for every link $e\in\mathcal{E}_{\mathrm{inf}}$, where $\mathcal{D}_e^\mu$ is a refinement of $\prod_{e\in\mathcal{E}}[\underline{x}_e^\mu, \bar{x}_e^\mu]$ by using monotone network dynamics. Recalling the definition of $N_e(s,x)$ and the physical meanings of $I_e(s,x)$ and $O_e(s,x)$, we see that the lower bounds $\underline{x}_m^\mu$ and the upper bounds $\bar{x}_n^\mu$ in $\mathcal{D}_e^\mu$ make our stability verification consider as many flows as possible from the origin into link $e$ to act as the inflow of link $e$. Note that $x_n$ will take the infinite value if $\bar{x}_n^\mu=\infty$; we conclude the limit exists as $x_n\to\infty$ by noting that $N_e(s,x)$ is non-decreasing in $x_n$ for $n\in\mathcal{E}\setminus\{\mathcal{A}_e^-\cup\{e\}\cup\mathcal{B}_e^+\}$ (see the proof in Section III-C) and that $N_e(s,x)$ is bounded.

Another observation on Theorem 1 is that we use a rectangle-like set $\prod_{e\in\mathcal{E}}[\underline{x}_e^\mu,\bar{x}_e^\mu]$, instead of $\mathcal{X}^\mu$, to simplify the computation.  Though simple, such kind of set suffices to yield interesting results; see Theorems~\ref{thm_infinite} and \ref{thm_observable}. In general, the invariant set $\mathcal{X}^\mu$ could be in other shapes; the stability condition can be further refined given a specific invariant set.

One can obtain a lower bound of the resiliency score by finding the supremum of those demands $\alpha$ that satisfy the criterion \eqref{eq_psG}. It involves solving a set of maximization problems only over compact sets $\prod_{\ell\in\mathcal{B}_e^+}[\underline{x}_\ell^\mu,\bar{x}_\ell^\mu]$. Thus we are able to find the global optimal solution efficiently with searching algorithms. This is a significant refinement with respect to the general stability criteria, which essentially require search over unbounded sets \cite{benaim15}. Note that though the criterion \eqref{eq_psG} is a sufficient condition, it is also necessary in particular settings; see Theorems 3 and 4. 

Theorem~\ref{thm_stable} is proved based on a novel Lyapunov function \eqref{eq_Ly1} relying on the weight function \eqref{eq_rho}. We enhance it by considering more general weights for links from the sets $\mathcal{A}_{e,\mathrm{fin}}^-$ and $\mathcal{B}_e^+$, which is achieved by a more sophisticated Lyapunov function \eqref{eq_generalized_lyapunov}. Weighting link dynamics $G_\ell(s,x)$,  $\ell\in\mathcal{A}_{e,\mathrm{fin}}^-$, can enhance the stability analysis if there are upstream bottlenecks with congestion spillback. Note that by appropriately restricting the weights for $\mathcal{A}_{e,\mathrm{fin}}^-$ and $\mathcal{B}_e^+$, we only need to consider $x_m=\underline{x}_m^\mu$, $m\in\mathcal{A}_{e,\mathrm{inf}}^-$, when checking the stability, just like Theorem~1. We do not consider weighting $G_m(s,x)$,  $m\in\mathcal{A}_{e,\mathrm{inf}}^-$ for two reasons. First, link $m\in\mathcal{A}_{e,\mathrm{inf}}^-$ has infinite storage and does not block its inflow. Second, weighting $G_m(s,x)$, $m\in\mathcal{A}_{e,\mathrm{inf}}^-$, will destroy the monotonicity with respect to $x_m$, which incurs additional computational costs of the stability verification. 

For $
\mathcal{A}_{e,\mathrm{fin}}^-\cup\mathcal{B}_e^+=\{e_{i_1},e_{i_2},\cdots,e_{i_H}\}$, we define its state vector by $$x_e^*:=[x_{e_{i_1}},x_{e_{i_2}},\cdots,x_{e_{i_H}}]^{\mathrm{T}}\in\prod_{h=1}^H \mathcal{X}_{e_{i_h}}^\mu.$$ 
Along with the advanced Lyapunov function \eqref{eq_generalized_lyapunov}, we let
\begin{align}
    N_e^{*}(s, x)
    :=& \sum_{m\in\mathcal A_{e,\mathrm{inf}}^{-}\cup\{e\}} G_{m}(s,x) + 2 \xi_k^{\mathrm{T}}(x_e^{*})
    B_{se}
    \dot{\xi}_k(x_e^*) \nonumber \\
    &+ \sum_{s'\in\mathcal{S}} \lambda_{ss'}
    \xi_k^{\mathrm{T}}(x_e^*)
    (B_{s'e}-B_{se})
    \xi_k(x_e^*) \label{eq_netflow_plus}
\end{align}
denote a generalization of $N_e(s,x)$, where 
$$\xi_k(x_e^*):=[1,x_{e_{i_1}},\cdots,x_{e_{i_H}},\cdots,x_{e_{i_1}}^k,\cdots,x_{e_{i_H}}^k]^{\mathrm{T}}\in\mathbb{R}_{\geq0}^{kH+1}$$ is a monomial basis of degree $k$ for $x_e^*$,  
$$\dot{\xi}_k(x_e^{*}):=\frac{\mathrm{d}}{\mathrm{d}t} \xi_k(x_e^{*}(t))$$
is the derivative of $\xi_k(x_e^{*})$ with respect to time $t$, and $B_{se}\in\mathbb{S}_+^{kH+1}$ is a symmetric positive definite matrix for $s\in\mathcal{S}$ and $e\in\mathcal{E}_{\mathrm{inf}}$.

The second main result of this section is as follows:
\begin{thm} \label{thm_stable_stronger}
Consider an acyclic network satisfying Assumptions~\ref{asm_fault}.1-\ref{asm_fault}.3. Suppose that the network admits a demand $\alpha$ and  an invariant set $\mathcal X^\mu\subseteq\mathcal\prod_{e\in\mathcal{E}}[\underline{x}_e^\mu, \bar{x}_e^\mu]$ under a control $\mu:\mathcal S\times\mathcal X\to\mathbb R_{\ge0}^{\mathcal{P}}$. Besides, there exist a set of symmetric positive definite matrices $B_{se}\in\mathbb{S}_+^{kH+1}$, $e\in\mathcal{E}_{\mathrm{inf}}$, $s\in\mathcal{S}$, such that
\begin{equation}
    \bm{0}\leq 2J_{\xi_k}^{\mathrm{T}}(x_e^*)B_{se}\xi_k(x_e^*) \leq \bm{1},~\forall   x_e^{*}\in\prod_{h=1}^H\mathcal{X}_{e_{i_h}}^{\mu}
    \label{eq_Bse_restrict},
\end{equation}
where 
\begin{equation*}
    J_{\xi_k}(x_e^*):=
\frac{\mathrm{d}}{\mathrm{d}x_e^*}\xi_k(x_e^{*})\in\mathbb{R}^{(kH+1)\times H} 
\end{equation*}
is a Jacobian matrix. Then, the network is stable if
    \begin{equation}
        \max_{x\in\mathcal{D}^{\mu*}_e} N_e^{*}(s,x) < 0,~\forall e\in\mathcal{E}_{\mathrm{inf}},s\in\mathcal{S},
    \end{equation}
    where  $\mathcal{D}_{e}^{\mu*}$ is given by
    \begin{align}
       \mathcal{D}_e^{\mu*}:=\{x|&x_e=x_e^{c*},~x_m=\underline{x}_m^\mu,~\forall m \in \mathcal{A}_{e,\mathrm{inf}}^{-}, \nonumber \\
        &x_n=\bar{x}_n^\mu,~\forall n\in\mathcal{E}\setminus(\mathcal{A}^-_{e}\cup\{e\}\cup\mathcal{B}_e^+), \nonumber \\
        & x_\ell\in[\underline{x}_\ell^\mu,\bar{x}_\ell^\mu], ~\forall \ell\in\mathcal{A}_{e,\mathrm{fin}}^-\cup\mathcal{B}_e^+\}. \label{eq_compact_D_general}
    \end{align}
\end{thm}

The weight constraint \eqref{eq_Bse_restrict} induces each weight function of the link dynamics $G_{e_{i_h}}(s,x)$ to have a range $[0, 1]$. The non-negativity endows the weighted $G_{e_{i_h}}(s,x)$ with the correct physical meaning of net flows, and less than one makes $x_m$, $m\in\mathcal{A}_{e,\mathrm{inf}}^-\subseteq\mathcal{A}_e^-$, attain the minimum value by monotone network dynamics, like Theorem~\ref{thm_general}.

The proof of Theorem~\ref{thm_stable_stronger} is based on the generalized Lyapunov function \eqref{eq_generalized_lyapunov}. Note that the extension is not unique. For example, we can further adapt the Lyapunov function if the control law is specified; see our proof of Theorem 3 in Section~\ref{sec_pf_thm3_suf}. 

In general, Theorem~\ref{thm_stable_stronger} yields stronger criteria than Theorem~\ref{thm_stable}. The throughput bounds derived from Theorem~\ref{thm_stable_stronger} can be further sharpened by increasing the degree $k$ of the basis $\xi_k(x_e^*)$, but it requires more computational costs. We check the stability by solving the following \emph{semi-infinite programming} (SIP \cite{stein2012solve}) with finite decision variables but infinite constraints:
\begin{align}
        (\mathrm{P}_2)~&\min_{\gamma, B_{se}}\gamma \nonumber \\
        s.t.~& \eqref{eq_Bse_restrict} \nonumber ,\\
        &\gamma \geq N_e^{*}(s,x), ~\forall (s,x)\in\mathcal{S}\times\mathcal{D}_e^{\mu*}, \forall e\in\mathcal{E}_{\mathrm{inf}}. \label{eq_con}
\end{align}

The programming problem has infinite constraints because \eqref{eq_Bse_restrict} and \eqref{eq_con} should hold for infinitely many states $x$. If the optimal $\gamma^*<0$, we say the network is stable. Since link $\ell\in\mathcal{A}_{e,\mathrm{fin}}^-\cup\mathcal{B}_e^+$ is density-bounded, the constraint \eqref{eq_Bse_restrict} is required to hold over the compact set $\prod_{h=1}^H\mathcal{X}_{e_{i_h}}^{\mu}$ as well. Thus we can solve $\mathrm{P}_2$ efficiently with the solution algorithms for SIPs, such as adaptive convexification \cite{stein2012solve}.

The rest of this section is devoted to a numerical example for resiliency analysis based on the above results (Section~\ref{sub_exm1}) and the proof of Theorem~\ref{thm_stable} (Section~\ref{sub_thm1}) and Theorem~\ref{thm_stable_stronger} (Section~\ref{sub_thm2}).

\subsection{Numerical example}
\label{sub_exm1}

Consider the network in Fig.~\ref{fig_network}. The sending flow function of link $e$ is given by $f_e(s,x_e)=\min\{v_fx_e,Q_{se}\}$, where $v_f>0$ is a coefficient of free-flow speed \cite{daganzo95} and $Q_{se}$ is the mode-specific capacity of link $e$. The receiving flow functions are given by
$$
r_e(x_e)=
\begin{cases}
\min\{w_c(x_e^{\max}-x_e), Q_e\}, & \text{if } x_e^{\max}<\infty, \\
\infty, & \text{if } x_e^{\max}=\infty,
\end{cases}
$$
where $w_c>0$ is a coefficient of congestion-wave speed \cite{daganzo95} and $Q_e$ is the nominal capacity. In this example, we set $v_f=1$, $w_c=1/2$, $Q_{e_o}=Q_{e_1}=Q_{e_5}=1$, and $Q_{e_2}=Q_{e_3}=Q_{e_4}=1/2$. If links $e_1,\cdots,e_5$ have finite storage space, we let $x_{e_1}^{\max}=x_{e_5}^{\max}=3$, and $x_{e_2}^{\max}=x_{e_3}^{\max}=x_{e_4}^{\max}=3/2$.

The network is subject to cyber-physical disruptions as follows. On link $e_4$, the traffic state $x_{e_4}$ may temporarily appear to be zero to the controller.
On link $e_5$, the sending capacity can be temporarily reduced to zero.
Table~\ref{tab_modes} characterizes the fault mapping $T_{e_4}(s,x)$ for cyber disruptions and the mode-specific capacity $Q_{se_5}$ for physical disruptions. The other links are not subject to disruptions; i.e. $T_{e}(s,x)=x_e$ for all $s$ and all $e\neq e_4$, and $Q_{se}=Q_e$ for all $s$ and all $e\neq e_5$. Hence, the network has at most four modes.

\begin{table}[hbt]
\centering
\caption{Modes of network in Figure~\ref{fig_network}.}
\label{tab_modes}
\small
\begin{tabu}to0.6\linewidth{X[2,c,m]X[c,m]X[c,m]X[c,m]X[c,m]}
\toprule
Mode & $s_0$ & $s_1$ & $s_2$ & $s_3$ \\ \midrule
$T_{e_4}(s,x)$    & $x_{e_4}$  & $x_{e_4}$ & $0$ & $0$  \\
$Q_{se_5}$     & $1$ & $0$   & $1$ &  $0$ \\
\bottomrule
\end{tabu}
\end{table}

If the network suffers the cyber and physical disruption simultaneously, the mode transition rate matrix is
\begin{equation*}
\Lambda =
\left[\begin{array}{cccc}
     -0.2 & 0.1 & 0.1 & 0 \\
     0.1 & -0.2 & 0 & 0.1 \\
     0.1 & 0 & -0.2 & 0.1 \\
     0 & 0.1 & 0.1 & -0.2
\end{array}
\right].
\end{equation*}
The corresponding steady-state probabilities $p_s$ can be computed for the mode $s\in\{s_0, s_1, s_2, s_3\}$: $p_{s_0}=p_{s_1}=p_{s_2}=p_{s_3}=1/4$.

In the following we specify network dynamics. The traffic flows at the diverges are routed according to the classical logit model:
\begin{align*}
    &q_{e_oe_1}^{\mathrm{logit}}(s,x)=\min\{\frac{e^{-\nu x_{e_1}}}{e^{-\nu x_{e_1}}+e^{-\nu T_{e_4}(s,x)}}f_{e_o}(s,x_{e_o}),r_{e_1}(x_{e_1})\},\\
    &q_{e_oe_4}^{\mathrm{logit}}(s,x)=\min\{\frac{e^{-\nu T_{e_4}(s,x)}}{e^{-\nu x_{e_1}}+e^{-\nu T_{e_4}(s,x)}}f_{e_o}(s,x_{e_o}),r_{e_4}(x_{e_4})\},\\
    &q_{e_1e_2}^{\mathrm{logit}}(s,x)=\min\{\frac{e^{-\nu x_{e_2}}}{e^{-\nu x_{e_2}}+e^{-\nu x_{e_3}}}f_{e_1}(s,x_{e_1}),r_{e_2}(x_{e_2})\},\\
    &q_{e_1e_3}^{\mathrm{logit}}(s,x)=\min\{\frac{e^{-\nu x_{e_3}}}{e^{-\nu x_{e_2}}+e^{-\nu x_{e_3}}}f_{e_1}(s,x_{e_1}),r_{e_3}(x_{e_3})\}.
\end{align*}
In this example, we select $\nu=2$ as the sensitivity coefficient.
At the merge, we assume that link $e_2$ is prioritized over link $e_4$.
Such a priority can be realized by
\begin{align*}
    &q_{e_2e_5}^{\mathrm{logit}}(s,x)=\min\{f_{e_2}(s,x_{e_2}),r_{e_5}(x_{e_5})\},\\
    &q_{e_4e_5}^{\mathrm{logit}}(s,x)=\min\{f_{e_4}(s,x_{e_4}),r_{e_5}(x_{e_5})-q_{e_2e_5}^{\mathrm{logit}}(s,x)\}.
\end{align*}


\begin{table}[htbp]
    \centering
    \scriptsize
    \caption{Resiliency scores in various scenarios.}
    \begin{tabu}to\linewidth{X[1,c,m]X[2,c,m]X[2,c,m]X[c,m]X[c,m]X[c,m]}
        \toprule
        Infinite space & Cyber disruptions & Physical disruptions & $\underline{\eta}^{\mathrm{logit}}_1$ &  $\underline{\eta}^{\mathrm{logit}}_2$ &  $\eta^{\mathrm{logit}}_{\mathrm{sim}}$ \\
        \midrule
        yes & no & no & 1 & 1 & 1 \\
        yes & yes & no & 0.783 & 0.984 & 1 \\
        yes & no & yes & 0.667  & 0.667 & 0.667 \\
        yes & yes & yes & 0.602 & 0.623  & 0.637 \\
        no & no & no & 1 & 1 & 1 \\
        no & yes & no & 0.792 & 0.910 & 0.921 \\
        no & no & yes & 0.741 & 0.848 & 0.860 \\
        no & yes & yes & 0.493 & 0.708 & 0.722 \\
        \bottomrule
    \end{tabu}
    \label{tab_analysis}
\end{table}
Table~\ref{tab_analysis} lists resiliency scores in various scenarios. The column ``Infinite space'' indicates whether links $e_1,e_2,\cdots,e_5$ have infinite storage space. Clearly, we have the nominal MCC equal to one, i.e. $C(\mathcal{Q}_{s_0};\mathcal{G})=1$. The lower bounds $\underline{\eta}^{\mathrm{logit}}_1$ are derived from Theorem~\ref{thm_stable}, while the bounds $\underline{\eta}^{\mathrm{logit}}_2$ are obtained through Theorem~\ref{thm_stable_stronger}. The details are available in the supplementary material. We also used numerical simulation to evaluate true resiliency scores, denoted by $\eta^{\mathrm{logit}}_{\mathrm{sim}}$. Our findings are summarized below, where the finding (iii) is a little surprising at the first glance:
\begin{enumerate}[(i)]
    \item  It is demonstrated that Theorem~\ref{thm_stable_stronger} yields tighter lower bounds than Theorem~\ref{thm_stable}. More importantly, the bounds $\underline{\eta}_2^{\mathrm{logit}}$ are close to $\eta_{\mathrm{sim}}^{\mathrm{logit}}$, i.e. the resiliency scores revealed by numerical simulation.
    \item Cyber-physical disruptions can significantly decrease the network performance. For instance, the resiliency score decreases to 0.860 when links $e_1,\cdots,e_5$ only have finite space and the network suffers the physical disruption. It further declines to 0.722 when the network is also subject to the cyber disruption.
    \item Increasing link storage may not help improve resiliency under inappropriate localized controls. As seen in Table~\ref{tab_analysis}, the disruptions have more significant impacts given infinite link storage space. The reason lies in that the congestion in link $e_5$ does not affect the upstream routing when link $e_5$ has infinite storage.
\end{enumerate}

\subsection{Proof of Theorem~\ref{thm_stable}}
\label{sub_thm1}

The proof consists of two steps. We first prove Proposition~\ref{prp_stable}, and then show that under Assumption~\ref{asm_fault}.3 we only need to check over $\mathcal{D}_e^\mu$ given by \eqref{eq_compact_D} instead of the unbounded set $\tilde{\mathcal{D}}_e^\mu$ in Proposition~\ref{prp_stable}. 
\begin{prp}
\label{prp_stable}
Consider an acyclic network satisfying Assumptions~\ref{asm_fault}.1-\ref{asm_fault}.2. Suppose that the network admits a demand $\alpha$ and an invariant set $\mathcal X^\mu\subseteq\prod_{e\in\mathcal{E}}[\underline{x}_e^\mu, \bar{x}_e^\mu]$ under a control law $\mu:\mathcal S\times\mathcal X\to\mathbb R_{\ge0}^{\mathcal{P}}$. Then the network is stable if
    \begin{equation}
        \sum_{s\in\mathcal S}p_s\max_{x\in\tilde{\mathcal{D}}^\mu_e}N_e(s,x)<0,~\forall e\in \mathcal{E}_{\mathrm{inf}}, \label{eq_psG_prp}
    \end{equation}
    where $\tilde{\mathcal{D}}_e^\mu:=\{x\in\mathcal{X}^\mu|x_e\geq x_e^{c*}\}$.
\end{prp}

We prove Proposition~\ref{prp_stable} by applying the following Foster-Lyapunov criterion adapted from Theorem~4.3 (ii) in \cite{meyn93}:
\begin{namedthm*}{Foster-Lyapunov criterion}
Consider a non-explosive piecewise-deterministic Markov process $\{(S(t),X(t)); t\ge0\}$ with a state space $\mathcal{S}\times\mathcal X$, an infinitesimal generator $\mathscr L$, and a Lyapunov function $V:\mathcal{S}\times\mathcal X\to\mathbb R_{\ge0}$. If there exist constants $c>0$, $d<\infty$ and a function $f:\mathcal X\to\mathbb R_{\ge0}$, with $f(x)\to\infty$ as $|x|\to\infty$, such that
\begin{align*}
    \mathscr LV(s,x)\le-cf(x)+d,
    ~\forall (s,x)\in\mathcal{S}\times\mathcal X,
\end{align*}
then, for each initial condition $(s,x)\in\mathcal{S}\times\mathcal X$,
$$
\limsup_{t\to\infty}\frac1t\int_{\tau=0}^t \mathbb{E}[f(X(\tau))|S(0)=s,X(0)=x]\mathrm{d}\tau\le d/c.
$$
\end{namedthm*}

Consider a novel switched polynomial Lyapunov function:
\begin{align}
    V(s,x) :=& \sum_{e\in\mathcal{E}_{\mathrm{inf}}} a_e x_{e}\Big[\frac{1}{2}x_e+\sum_{m\in\mathcal{A}_{e}^{-}} x_m + \sum_{\ell\in\mathcal{B}^{+}_{e}}\int^{x_{\ell}}_{\underline{x}_{\ell}}\rho_{\ell}(\zeta)\mathrm{d}\zeta + b_{se}\Big] \label{eq_Ly1}
\end{align}
where $a_e>0$ and $b_{se}\geq0$ for any $s\in\mathcal{S}$ and $e\in\mathcal{E}_{\mathrm{inf}}$. The design of \eqref{eq_Ly1} is motivated by network dynamics:
\begin{enumerate}[(i)]
    \item For each $e\in\mathcal{E}_{\mathrm{inf}}$, its upstream links, $\mathcal{A}_e^-$, and downstream bottlenecks, $\mathcal{B}_e^+$, are taken into account;
    \item The weight function \eqref{eq_rho} is introduced to better capture bottleneck dynamics. If $\rho_\ell\equiv1$, \eqref{eq_Ly1} is reduced to a piecewise quadratic function.
    \item The parameters $a_e$ and $b_{se}$ are used to model the impact of mode $s$ on link $e$. We will show that such a design saves finding explicit values of $a_e$ and $b_{se}$.
\end{enumerate}

In the following we show that if \eqref{eq_psG_prp} holds, we must have $a_{e}>0$, $b_{se}\geq0$, $c>0$ and $d<\infty$ such that $V(s,x)$ satisfies
\begin{equation}
    \mathscr L V(s,x) \leq -c\sum_{e\in\mathcal{E}_{\mathrm{inf}}} x_e + d, ~\forall (s,x)\in\mathcal{S}\times\mathcal{X}^\mu. \label{eq_drift_cond}
\end{equation}

To proceed, we apply the infinitesimal generator $\mathscr L$ to the Lyapunov function $V(s,x)$. By \eqref{eq_infinitesimal}, we have
\begin{align*}
    \mathscr{L}V(s,x) 
    &= \sum_{e\in\mathcal{E}_{\mathrm{inf}}} \bigg(a_ex_e\Big(N_e(s,x) + \lambda_s^{\mathrm{T}}(b_e-b_{se}\mathbf{1})\Big) \\
    & \quad + a_eG_e(s,x)\Big(\sum_{m\in\mathcal{A}_e^-}x_m + \sum_{\ell\in\mathcal{B}_e^+}\int_{\underline{x}_\ell}^{x_\ell}\rho(\zeta)\mathrm{d}\zeta + b_{se}\Big)\bigg),
\end{align*}
where $\bm{1}$ denotes the all-ones vector, $\lambda_s:= [\lambda_{ss_0},\cdots,\lambda_{ss_M}]^{\mathrm{T}}$ and $b_e:=[b_{s_0e},\cdots,b_{s_Me}]^{\mathrm{T}}$.

Noting
\begin{align*}
    \sum_{e\in\mathcal{E}_{\mathrm{inf}}} G_e(s,x)\sum_{m\in\mathcal{A}^{-}_{e,\mathrm{inf}}}x_m
    =& \sum_{\substack{e\in\mathcal{E}_{\mathrm{inf}} \\ m\in\mathcal{E}_{\mathrm{inf}}}} G_e(s,x)x_m \mathbb{I}_{\mathcal{A}^{-}_{e,\mathrm{inf}}}(m) \\
    =& \sum_{\substack{e\in\mathcal{E}_{\mathrm{inf}} \\ n\in\mathcal{E}_{\mathrm{inf}}}} G_n(s,x)x_e \mathbb{I}_{\mathcal{A}^{+}_{n,\mathrm{inf}}}(e) \\
    =&\sum\limits_{e\in\mathcal{E}_{\mathrm{inf}}} x_e\sum\limits_{n\in\mathcal{A}_{e,\mathrm{inf}}^{+}} G_n(s,x),
\end{align*}
we obtain
\begin{align*}
    \mathscr L V(s,x) =&
    \sum_{
    e\in\mathcal{E}_{\mathrm{inf}}}
    \bigg(D_{e}(s,x)x_{e} +a_eG_{e}(s,x)\Big(\sum_{m\in\mathcal{A}_{e,\mathrm{fin}}^-} x_m + \sum_{\ell\in\mathcal{B}^+_{e}} \int_{\underline{x}_\ell}^{x_\ell} \rho_{\ell}(\zeta)\mathrm{d}\zeta + b_{se}\Big)\bigg),
\end{align*}
where $\mathbb{I}_{\mathcal{A}}(e)$ is an indicator function such that $\mathbb{I}_{\mathcal{A}}(e)=1$ if $e\in\mathcal{A}$ and $\mathbb{I}_{\mathcal{A}}(e)=0$ otherwise,
and
\begin{align*}
    D_e(s,x) &:=  a_e\Big(N_{e}(s,x) + \lambda_s^{\mathrm{T}}(b_e - b_{se}\bm{1})\Big)  + \sum\limits_{n\in\mathcal{A}_{e,\mathrm{inf}}^{+}} a_n G_n(s,x).
\end{align*}
 It follows 
$\mathscr{L}V(s,x)\leq \sum_{e\in\mathcal{E}_{\mathrm{inf}}} D_e(s,x)x_e + d_0$ with $d_0<\infty$. The existence of $d_0$ is guaranteed by the boundedness of $x_m$ for $m\in\mathcal{A}_{e,\mathrm{fin}}^-$, of $x_\ell$ for $\ell\in\mathcal{B}^+_{e}$ and of $G_e(s,x)$ for $e\in\mathcal{E}$.

Next, we show, if \eqref{eq_psG_prp} holds, there exist $a_{e} > 0$, $b_{se} \geq 0$, $c_e>0$ and $d_e < \infty$ such that
\begin{equation}
D_{e}(s, x)x_{e} \le -c_{e} x_{e} + d_{e}, ~\forall (s, x)\in\mathcal{S}\times\mathcal{X}^\mu,\forall e\in\mathcal{E}_{\mathrm{inf}},
\label{eq_drift}
\end{equation}
which implies that \eqref{eq_drift_cond} is satisfied with $c:= \min_{e\in\mathcal{E}_{\mathrm{inf}}}c_e$ and $d
:=\sum_{e\in\mathcal{E}_{\mathrm{inf}}}d_e + d_0$.

For any $e\in\mathcal{E}_{\mathrm{inf}}$, we must have $ d_{e,1} < \infty$ such that
\begin{equation}
    \max_{(s,x)\in\mathcal{S}\times(\mathcal{X}^\mu\setminus\tilde{\mathcal{D}}_e^\mu)} D_e(s,x)x_e \leq d_{e,1}. \label{eq_prp_1}
\end{equation}
Besides, for any $(s,x)\in\mathcal{S}\times\tilde{\mathcal{D}}^\mu_e$,
\begin{align}
    D_e(s,x) \leq& a_e[\bar{N}_{se} + \lambda_s^{\mathrm{T}}(b_e-b_{se}\bm{1})] + \bar{G} \sum_{n\in\mathcal{A}_{e,\mathrm{inf}}^{+}} a_n \nonumber \\
    \leq& a_e\sum_{s'\in\mathcal{S}}p_{s'}\bar{N}_{s'e} + \bar{G} \sum_{n\in\mathcal{A}_{e,\mathrm{inf}}^{+}} a_{n} \label{eq_pf_prp1_1}
\end{align}
where $\bar{N}_{se}:=\max_{x\in\tilde{\mathcal{D}}_e^\mu} N_e(s,x)$ and $\bar{G}$ denotes an upper bound of $G_e(s,x)$ for $e\in\mathcal{E}$. The last step is derived from the following lemma, which we prove in Appendix~\ref{app_lmm1}.
\begin{lmm}
\label{lmm_1}
The system of equations 
\begin{equation}
    z_{s}+\sum_{s'\in\mathcal{S}}\lambda_{ss'}(y_{s'}-y_{s})=\sum_{s'\in\mathcal S}p_{s'}z_{s'},
    ~s\in\mathcal{S} \label{eq_bks}
\end{equation}
has a non-negative solution for $y=[y_{s_0},\cdots,y_{s_M}]^{\mathrm{T}}$, where $p_s$, $\lambda_{ss'}$ and $z_s$ are given, and $p_s$ and $\lambda_{ss'}$ satisfy \eqref{eq_p}.
\end{lmm}

Given \eqref{eq_psG_prp} and \eqref{eq_pf_prp1_1}, Lemma 2 proved in Appendix~\ref{app_lmm2} tells that there exists $a_e$ for $e\in\mathcal{E}_{\mathrm{inf}}$ such that $D_e(s,x)<0,~\forall (s,x)\in\mathcal{S}\times\tilde{\mathcal{D}}_e^\mu,\forall e\in\mathcal{E}_{\mathrm{inf}}$. 
Thus we must have $c_e>0$ for each $e\in\mathcal{E}_{\mathrm{inf}}$ such that
\begin{equation}
    D_e(s,x)x_e \leq -c_ex_e, ~\forall (s,x)\in\mathcal{S}\times\tilde{\mathcal{D}}_e^\mu,\forall e\in\mathcal{E}_{\mathrm{inf}}. \label{eq_pf_prp2}
\end{equation}
\begin{lmm}\label{lmm_ak}
The system of equations
\begin{align}
    \pi_{e} a_e + \bar{G} \sum_{n\in\mathcal{A}_{e,\mathrm{inf}}^{+}}a_n < 0, ~\forall e\in{\mathcal E}_{\mathrm{inf}}
    \label{eq_ak}
\end{align}
has a positive solution for $\{a_e|e\in\mathcal{E}_{\mathrm{inf}}\}$ if $\pi_e<0$ for any $e\in\mathcal{E}_{\mathrm{inf}}$.
\end{lmm}

Combining \eqref{eq_prp_1} and \eqref{eq_pf_prp2} yields \eqref{eq_drift} with $d_e=d_{e,1}+c_ex_e^{c*}$.
It indicates that we finish the proof of Proposition~\ref{prp_stable}.

Noting $\mathcal{X}^\mu\subseteq\prod_{e\in\mathcal{E}}[\underline{x}_e^\mu, \bar{x}_e^\mu]$, we must have
$$
\max_{x\in\prod_{e\in\mathcal{E}}[\underline{x}_e^\mu, \bar{x}_e^\mu],x_e\geq x_e^{c*}} N_e(s,x) \geq  \max_{x\in\tilde{\mathcal{D}}_e^\mu} N_e(s,x).
$$
Then, the rest of the proof is devoted to showing that
\begin{equation}
    \max_{x\in\mathcal{D}_e^\mu} N_e(s,x) = \max_{x\in\prod_{e\in\mathcal{E}}[\underline{x}_e^\mu, \bar{x}_e^\mu],x_e\geq x_e^{c*}} N_e(s,x) \label{eq_reduction}
\end{equation}
under Assumption~\ref{asm_fault}.3. Using \eqref{eq_reduction}, we can derive \eqref{eq_psG_prp} from \eqref{eq_psG}, which finishes the proof. 

The definition of localized control laws indicates
\begin{equation*}
     \max_{x_{e'}} N_e(s,x)
    = \max_{x_{e'}}G_{e,\tau_{e'}}(s,x) + G_{e,\sigma_{e'}}(s,x),~\forall e'\in\mathcal{E},
\end{equation*}
where $G_{e,\tau_{e'}}(s,x)$ and $G_{e,\sigma_{e'}}(s,x)$ denote the projections of $N_e(s,x)$ onto nodes $\tau_{e'}$ and $\sigma_{e'}$, respectively. The projection onto an arbitrary node $v\in\mathcal{V}$ is given by 
\begin{align*}
    G_{e,v}(s,x) &:=\sum_{i\in\mathcal{E}_v^-,j_1\in\mathcal{E}_{e,v}^{1+}} q_{ij_1}(s,x) 
    +\sum_{i\in\mathcal{E}_v^-,j_1\in\mathcal{E}_{e,v}^{3+}} \rho_{j_3}(x_{j_3}) q_{ij_3}(s,x) \\
    &-\sum_{i_1\in\mathcal{E}_{e,v}^{1-},j\in\mathcal{E}_v^+} q_{i_1j}(s,x)
    - \sum_{i_3\in\mathcal{E}_{e,v}^{3-},j\in\mathcal{E}_v^+} \rho_{i_3}(x_{i_3}) q_{i_3j}(s,x),
\end{align*}
where $\rho_{j_3}(x_{j_3})$ and $\rho_{i_3}(x_{i_3})$ are specified by \eqref{eq_rho}, and the link sets $\mathcal{E}_{e,v}^{1+}$, $\mathcal{E}_{e,v}^{3+}$, $\mathcal{E}_{e,v}^{1-}$ and $\mathcal{E}_{e,v}^{3-}$ are illustrated in Fig.~\ref{fig_G_ev}(a). Note that these sets could be empty. We also use the example in Fig.~\ref{fig_network} to present two specific cases in Figs.~\ref{fig_G_ev}(b) and (c), respectively.

\begin{figure}[htbp]
    \begin{subfigure}{\linewidth}
    \centering
    \resizebox{0.6\linewidth}{!}{
    \begin{tikzpicture}
        \Vertex[x=0,size=0.1,style={black},label=$v$,position={below}]{D}
        
        \Vertex[x=-1,y=1,size=0.1,style={white}]{B}
        \node at (-2.2, 0.8) (B1) {\scriptsize $\underbrace{\mathcal{E}_{v}^-\cap(\mathcal{A}_{e}^-\cup\{e\})}_{i_1\in\mathcal{E}_{e,v}^{1-}}$};
        
        \Vertex[x=-1.5, y=0, size=0.1, style={white}]{C}
        \node at (-3, -0.2) (C1) {\scriptsize $\underbrace{\mathcal{E}_v^-\setminus(\mathcal{A}_e^-\cup\{e\}\cup\mathcal{B}_e^+)}_{i_2\in\mathcal{E}_{e,v}^{2-}}$};
        
        \Vertex[x=-1,y=-1,size=0.1,style={white}]{A}
        \node at (-2.2, -1.2) (A1) {\scriptsize $\underbrace{\mathcal{E}_{v}^-\cap\mathcal{B}_e^+}_{i_3\in\mathcal{E}_{e,v}^{3-}}$};
        
        \Vertex[x=1, y=1, size=0.1,style={white}]{G}
        \node at (2.2, 0.8) (G1)
        {\scriptsize
        $\underbrace{\mathcal{E}_{v}^+\cap(\mathcal{A}^-_{e}\cup\{e\})}_{j_1\in\mathcal{E}_{e,v}^{1+}}$};
        
        \Vertex[x=1.5, y=0, size=0.1, style={white}]{E}
        \node at (3, -0.2) (E1) {\scriptsize $\underbrace{\mathcal{E}_v^+\setminus(\mathcal{A}_e^-\cup\{e\}\cup\mathcal{B}_e^+)}_{j_2\in\mathcal{E}_{e,v}^{2+}}$};

        \Vertex[x=1, y=-1,
        size=0.1,style={white}]{F}
        \node at (2.2, -1.2) (F1) {\scriptsize $\underbrace{\mathcal{E}_{v}^+\cap\mathcal{B}_e^+}_{j_3\in\mathcal{E}_{e,v}^{3+}}$};
        
        \Edge[Direct,lw=1pt](A)(D)
        \Edge[Direct,lw=1pt](B)(D)
        \Edge[Direct,style={dashed},lw=1pt](C)(D)
        \Edge[Direct,style={dashed},lw=1pt](D)(E)
        \Edge[Direct,lw=1pt](D)(F)
        \Edge[Direct,lw=1pt](D)(G)

        \Vertex[x=0,y=1,size=0.5,style={white}]{X}
        \Vertex[x=0,y=-1,size=0.5,style={white}]{Y}
        \Edge[style={dotted},bend=60](X)(Y)
        \Edge[style={dotted},bend=60](Y)(X)

        \Vertex[x=0.3,y=1.2,size=0.5,style={white},label=$j\in\mathcal{E}_{v}^+$]{XX}
        \Vertex[x=-0.3,y=-1.2,size=0.5,style={white},label=$i\in\mathcal{E}_{v}^-$]{YY}
        \end{tikzpicture}
        }
        \caption {}
        \end{subfigure}

    \begin{subfigure}{0.4\linewidth}
    \centering
    \resizebox{\linewidth}{!}{
    \begin{tikzpicture}
        \Vertex[x=-0.9,style={color=white}]{A}
        \Vertex[label=$v_o$,x=1,fontscale=1]{B}
        \Vertex[label=$v_1$,x=2.5,y=1,fontscale=1]{C}
        \Vertex[label=$v_2$,x=2.5,y=-1,fontscale=1]{D}
        \Vertex[label=$v_d$,x=4,fontscale=1]{E}
        \Edge[Direct,label=$e_o$,fontscale=1,style=dotted](A)(B)
        
        \Edge[Direct,label=$\mathcal{E}_{e_o,v_1}^{3-}$,fontscale=1,position=above left](B)(C)
        \draw[dashed] (1.8,0.9) to (2.1,0.4);
        \Edge[Direct,label=$\mathcal{E}_{e_o,v_1}^{3+}$,fontscale=1,position=left](C)(D)
        \draw[dashed] (2.2,0.2) to (2.8,0.2);
        \Edge[Direct,label=$\mathcal{E}_{e_o,v_1}^{2+}$,fontscale=1,position=above right](C)(E)
        \draw[dashed] (3.2,0.9) to (2.9,0.4);
        
        \Edge[Direct,label=$e_4$,fontscale=1](B)(D)
        \Edge[Direct,label=$e_5$,fontscale=1,style=dotted](D)(E)
        
    \end{tikzpicture}
    }
    \caption{ }
    \end{subfigure}
    \begin{subfigure}{0.4\linewidth}
    \centering
    \resizebox{\linewidth}{!}{
    \begin{tikzpicture}
        \Vertex[x=-0.9,style={color=white}]{A}
        \Vertex[label=$v_o$,x=1,fontscale=1]{B}
        \Vertex[label=$v_1$,x=2.5,y=1,fontscale=1]{C}
        \Vertex[label=$v_2$,x=2.5,y=-1,fontscale=1]{D}
        \Vertex[label=$v_d$,x=4,fontscale=1]{E}
        \Edge[Direct,label=$e_o$,fontscale=1,style=dotted](A)(B)

        \Edge[Direct,label=$\mathcal{E}_{e_5,v_1}^{1-}$,fontscale=1,position=above left](B)(C)
        \draw[dashed] (1.8,0.9) to (2.1,0.4);
        \Edge[Direct,label=$\mathcal{E}_{e_5,v_1}^{1+}$,fontscale=1,position=left](C)(D)
        \draw[dashed] (2.2,0.2) to (2.8,0.2);
        \Edge[Direct,label=$\mathcal{E}_{e_5,v_1}^{2+}$,fontscale=1,position=above right](C)(E)
        \draw[dashed] (3.2,0.9) to (2.9,0.4);
        
        \Edge[Direct,label=$e_4$,fontscale=1](B)(D)
        \Edge[Direct,label=$e_5$,fontscale=1,style=dotted](D)(E)
        
    \end{tikzpicture}
    }
    \caption{ }
    \end{subfigure}
    
    \caption{Classification of incoming links $\mathcal{E}_v^-$ and outgoing links $\mathcal{E}_v^+$ for $e\in\mathcal{E}_{\mathrm{inf}}$ and $v\in\mathcal{V}$: (a) a general case, (b) classification for link $e_o$ and node $v_1$, given $\mathcal{E}_{v_1}^-=\{e_1\}$, $\mathcal{E}_{v_1}^+=\{e_2, e_3\}$,  $\mathcal{A}_{e_o}^-=\varnothing$ and $\mathcal{B}_{e_o}^+=\{e_1,e_2\}$, (c) classification for link $e_5$ and node $v_1$, given $\mathcal{E}_{v_1}^-=\{e_1\}$, $\mathcal{E}_{v_1}^+=\{e_2, e_3\}$,
    $\mathcal{A}_{e_5}^-=\{e_o,e_1,e_2,e_4\}$ and $\mathcal{B}_{e_5}^+=\varnothing$.
    }
    \label{fig_G_ev}
\end{figure}

The following first shows $N_e(s,x)$ is non-decreasing in $x_n$ for $n\in\mathcal{E}\setminus(\mathcal{A}_e^-\cup\{e\}\cup\mathcal{B}_e^+)$ by proving $G_{e,\tau_n}(s,x)+G_{e,\sigma_n}(s,x)$ is non-decreasing in $x_n$. By Assumption 3.3, the incoming flow $\sum_{i\in\mathcal{E}_v^-}q_{ij}(s,x)$ into link $j\in\mathcal{E}_{e,v}^{1+}\cup\mathcal{E}_{e,v}^{3+}$ is non-decreasing in $x_{j_2}$; the outgoing flow $\sum_{j\in\mathcal{E}_v^+}q_{ij}(s,x)$ from link $i\in\mathcal{E}_{e,v}^{1-}\cup\mathcal{E}_{e,v}^{3-}$ is non-increasing in $x_{j_2}$. Thus $G_{e,v}(s,x)$ is non-decreasing in $x_{j_2}$. Similarly, it is easy to verify that $G_{e,v}(s,x)$ is non-decreasing in $x_{i_2}$. This implies that both  $G_{e,\sigma_n}(s,x)$ and $G_{e,\tau_n}(s,x)$ are non-decreasing in $x_n$ for any $n\in\mathcal{E}\setminus(\mathcal{A}_{e}^{-}\cup\{e\}\cup\mathcal{B}_{e}^{+})$.

Now we show $N_e(s,x)$ is non-increasing in $x_m$ for $m\in\mathcal{A}_e^-\cup\{e\}$ by proving $G_{e,\tau_m}(s,x)+G_{e,\sigma_m}(s,x)$ is non-increasing in $x_m$. Noting $\mathcal{E}_{e,v}^{1-}\cup\mathcal{E}_{e,v}^{2-}\cup\mathcal{E}_{e,v}^{3-} = \mathcal{E}_v^-$ and $\mathcal{E}_{e,v}^{1+}\cup\mathcal{E}_{e,v}^{2+}\cup\mathcal{E}_{e,v}^{3+} = \mathcal{E}_v^+$ from Fig.~\ref{fig_G_ev}, we rewrite $G_{e,v}(s,x)$ as 
\begin{align*}
    G_{e,v}(s,x) &= \sum_{i_2\in\mathcal{E}_{e,v}^{2-}, j\in\mathcal{E}_v^+} q_{i_2j}(s,x)
    + \sum_{i_3\in\mathcal{E}_{e,v}^{3-}, j\in\mathcal{E}_v^+}(1-\rho_{i_3}(x_{i_3}))q_{i_3j}(s,x) \\
    & - \sum_{i\in\mathcal{E}_v^-, j_2\in\mathcal{E}_{e,v}^{2+}} q_{ij_2}(s,x)
    - \sum_{i\in\mathcal{E}_v^-, j_3\in\mathcal{E}_{e,v}^{3+}}(1-\rho_{j_3}(x_{j_3}))q_{ij_3}(s,x).
\end{align*}
 By Assumption~3.3, the incoming flow $\sum_{i\in\mathcal{E}_v^-}q_{ij}(s,x)$ into link $j\in\mathcal{E}_{e,v}^{2+}\cup\mathcal{E}_{e,v}^{3+}$  is non-decreasing in $x_{j_1}$; the outgoing flow $\sum_{j\in\mathcal{E}_v^+}q_{ij}(s,x)$ from link $i\in\mathcal{E}_{e,v}^{2-}\cup\mathcal{E}_{e,v}^{3-}$ is non-increasing in $x_{j_1}$. Thus $G_{e,v}(s,x)$ is non-increasing in $x_{j_1}$. Similarly, we can show $G_{e,v}(s,x)$ is non-increasing in $x_{i_1}$. Thus we conclude both $G_{e,\sigma_m}(s,x)$ and $G_{e,\tau_m}(s,x)$ are non-increasing in $x_m$ for any $m\in\mathcal{A}_e^-\cup\{e\}$.

Thus we conclude \eqref{eq_reduction} by the monotonicity of $N_e(s,x)$, which completes the proof. \qed

\subsection{Proof of Theorem~\ref{thm_stable_stronger}}
\label{sub_thm2}

The proof is similar to that of Theorem~\ref{thm_stable}. The major difference lies in the following Lyapunov function:
\begin{align}
    V^*(s,x) :=& \sum_{
    e\in\mathcal{E}_{\mathrm{inf}}}
    a_ex_{e}\Big[\frac{1}{2}x_e + \sum_{m\in\mathcal{A}^{-}_{e,\mathrm{inf}}}x_m  +  \xi_k^{\mathrm{T}}(x_e^{*})B_{se} \xi_k(x_e^{*})\Big]. \label{eq_generalized_lyapunov}
\end{align}

Applying the infinitesimal generator yields
\begin{align*}
    \mathscr L V^*(s,x) \leq&
    \sum_{
    e\in\mathcal{E}_{\mathrm{inf}}}
    [D^{*}_{e}(s,x)x_{e} + d_0],
\end{align*}
where
$D_e^{*}(s,x) :=  a_e N_{e}^{*}(s,x)  + \sum_{n\in\mathcal{A}_{e,\mathrm{inf}}^{+}} a_n G_n(s,x)$.
Then we obtain the following result similar to Proposition~\ref{prp_stable}:
\begin{prp}
\label{prp_stable_general}
Consider an acyclic network satisfying Assumptions~\ref{asm_fault}.1-\ref{asm_fault}.2. Suppose that the network admits a demand $\alpha$ and an invariant set $\mathcal X^\mu\subseteq\prod_{e\in\mathcal{E}}[\underline{x}_e^\mu, \bar{x}_e^\mu]$ under a control law $\mu:\mathcal S\times\mathcal X\to\mathbb R_{\ge0}^{\mathcal{P}}$. Then the network is stable if
    \begin{equation}
        \max_{x\in\tilde{\mathcal{D}}^\mu_e}N_e^*(s,x)<0,~\forall e\in \mathcal{E}_{\mathrm{inf}},~\forall s\in\mathcal{S}, \label{eq_psG_prp_general}
    \end{equation}
    where $\tilde{\mathcal{D}}_e^\mu:=\{x\in\mathcal{X}^\mu|x_e\geq x_e^{c*}\}$.
\end{prp}

Because of $\mathcal{X}^\mu\subseteq\prod_{e\in\mathcal{E}}[\underline{x}_e^\mu, \bar{x}_e^\mu]$, we must have
$$
\max_{x\in\prod_{e\in\mathcal{E}}[\underline{x}_e^\mu, \bar{x}_e^\mu],x_e\geq x_e^{c*}} N_e^*(s,x) \geq  \max_{x\in\tilde{\mathcal{D}}_e^\mu} N_e^*(s,x).
$$
Next, we only need to prove
\begin{equation}
    \max_{x\in\mathcal{D}_{e}^{\mu*}} N_{e}^{*}(s,x) = \max_{x\in\prod_{e\in\mathcal{E}}[\underline{x}_e^\mu, \bar{x}_e^\mu],x_e\geq x_e^{c*}} N_e^*(s,x). \label{eq_pf_general}
\end{equation}
For $N_{e}^{*}(s,x)$ given by \eqref{eq_netflow_plus}, we notice
\begin{equation*}
    2\xi_k^{\mathrm{T}}(x_e^{*})
    B_{se}
    \dot{\xi}_k(x_e^{*}) = \underbrace{2
    \xi_k^{\mathrm{T}}(x_e^*)
    B_{se}J_{\xi_k}(x_e^*)
    }_{[\rho^{*}_{e_{i_1}}(x),\rho^{*}_{e_{i_2}}(x),\cdots,\rho^{*}_{e_{i_H}}(x)]}\begin{bmatrix}
    G_{e_{i_1}}(s,x) \\
    G_{e_{i_2}}(s,x)
    \\ \vdots \\ G_{e_{i_H}}(s,x)
    \end{bmatrix}.
\end{equation*}
Clearly, \eqref{eq_Bse_restrict} indicates $\rho^{*}_{e_{i_h}}(x)\in[0, 1]$ for $h=1,2,\cdots,H$. We see $\rho^{*}_{e_{i_h}}(x)$ as a generalization of $\rho_{e_{i_h}}(x_{e_{i_h}})$ given by \eqref{eq_rho}. 
\begin{figure}[htbp]
    \centering
    \begin{tikzpicture}
        \Vertex[x=0,size=0.1,style={black},label=$v$,position={below}]{D}
        
        \Vertex[x=-1,y=1,size=0.1,style={white}]{B}
        \node at (-2.2, 0.8) (B1) {\scriptsize $\underbrace{\mathcal{E}_{v}^-\cap(\mathcal{A}_{e,\mathrm{inf}}^-\cup\{e\})}_{i_1\in\mathcal{E}_{e,v}^{1-}}$};
        
        \Vertex[x=-1.5, y=0, size=0.1, style={white}]{C}
        \node at (-3, -0.2) (C1) {\scriptsize $\underbrace{\mathcal{E}_v^-\setminus(\mathcal{A}_e^-\cup\{e\}\cup\mathcal{B}_e^+)}_{i_2\in\mathcal{E}_{e,v}^{2-}}$};
        
        \Vertex[x=-1,y=-1,size=0.1,style={white}]{A}
        \node at (-2.2, -1.2) (A1) {\scriptsize $\underbrace{\mathcal{E}_{v}^-\cap(\mathcal{A}_{e,\mathrm{fin}}^-\cup \mathcal{B}_e^+ )}_{i_3\in\mathcal{E}_{e,v}^{3-}}$};
        
        \Vertex[x=1, y=1, size=0.1,style={white}]{G}
        \node at (2.2, 0.8) (G1)
        {\scriptsize
        $\underbrace{\mathcal{E}_{v}^+\cap(\mathcal{A}^-_{e,\mathrm{inf}}\cup\{e\})}_{j_1\in\mathcal{E}_{e,v}^{1+}}$};
        
        \Vertex[x=1.5, y=0, size=0.1, style={white}]{E}
        \node at (3, -0.2) (E1) {\scriptsize $\underbrace{\mathcal{E}_v^+\setminus(\mathcal{A}_e^-\cup\{e\}\cup\mathcal{B}_e^+)}_{j_2\in\mathcal{E}_{e,v}^{2+}}$};

        \Vertex[x=1, y=-1,
        size=0.1,style={white}]{F}
        \node at (2.2, -1.2) (F1) {\scriptsize $\underbrace{\mathcal{E}_{v}^+\cap(\mathcal{A}_{e,\mathrm{fin}}^-\cup \mathcal{B}_e^+)}_{j_3\in\mathcal{E}_{e,v}^{3+}}$};
        
        \Edge[Direct,lw=1pt](A)(D)
        \Edge[Direct,lw=1pt](B)(D)
        \Edge[Direct,style={dashed},lw=1pt](C)(D)
        \Edge[Direct,style={dashed},lw=1pt](D)(E)
        \Edge[Direct,lw=1pt](D)(F)
        \Edge[Direct,lw=1pt](D)(G)

        \Vertex[x=0,y=1,size=0.5,style={white}]{X}
        \Vertex[x=0,y=-1,size=0.5,style={white}]{Y}
        \Edge[style={dotted},bend=60](X)(Y)
        \Edge[style={dotted},bend=60](Y)(X)

        \Vertex[x=0.3,y=1.2,size=0.5,style={white},label=$j\in\mathcal{E}_{v}^+$]{XX}
        \Vertex[x=-0.3,y=-1.2,size=0.5,style={white},label=$i\in\mathcal{E}_{v}^-$]{YY}
    \end{tikzpicture}
    
    \caption{Classification of incoming links $\mathcal{E}_v^-$ and outgoing links $\mathcal{E}_v^+$ $e\in\mathcal{E}_{\mathrm{inf}}$ and $v\in\mathcal{V}$ in Theorem~2.
    }
    \label{fig_G_ev_2}
\end{figure}
Then we consider the link classification as shown in Fig.~\ref{fig_G_ev_2}. We can show the monotonicity of $N_e^*(s,x)$ with respect to i) $x_n$, $n\in\mathcal{E}\setminus(\mathcal{A}_e^-\cup\{e\}\cup\mathcal{B}_e^+)$, and ii) $x_m$, $m\in\mathcal{A}_{e,\mathrm{inf}}^-\cup\{e\}$, in a similar way to that in  Section~\ref{sub_thm1}, and thus prove \eqref{eq_pf_general}. \qed 
\section{Resilient control design}
\label{sec_control}

In this section, we study control design with guaranteed throughput. We will extend the classical notion of MCC to stochastic networks and present a series of results characterizing the resiliency under various classes of control laws.

\begin{figure}[hbtp]
    \centering
    \begin{subfigure}{0.32\linewidth}
    \includegraphics[width=\linewidth]{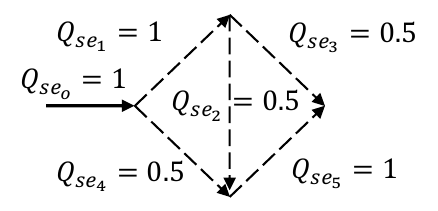}
        \caption{Min-cut with $C(\mathcal{Q}_s;\mathcal{G})=1$, $s\in\{s_0,s_2\}$.}
        \label{fig_min_cut_13}
    \end{subfigure}
    \begin{subfigure}{0.32\linewidth}
        \includegraphics[width=\linewidth]{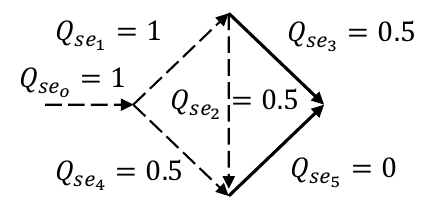}
        \caption{Min-cut with $C(\mathcal{Q}_s;\mathcal{G})=0.5$, $s\in\{s_1,s_3\}$.}
        \label{fig_min_cut_24}
    \end{subfigure}
    \begin{subfigure}{0.32\linewidth}
        \includegraphics[width=\linewidth]{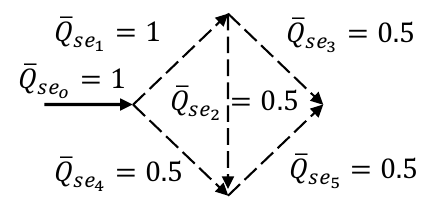}
        \caption{Min-cut with $C(\bar{\mathcal{Q}};\mathcal{G})=1$. $\quad\quad\quad\quad$}
        \label{fig_memm}
    \end{subfigure}
    \caption{EMCC and MECC; the link capacities are labeled, and the min-cuts are highlighted with solid links: a cut is a subset of links whose removal divides the node set into two disconnected subsets and min-cuts are those with the minimum sum of link capacities, i.e., the min-cut capacity.}
    \label{fig_cut}
\end{figure}

Two extensions of the MCC, i.e. the MECC and the EMCC, are considered in the stochastic setting.
The MECC is the minimum cut capacity evaluated with the expected capacity of each link, and the EMCC is the expected minimum cut capacity over various modes. Noting that $C(\mathcal{Q};\mathcal{G})$ denotes the MCC of network $\mathcal{G}$ with a set of link capacities $\mathcal{Q}$, we have
\begin{equation*}
    \mathrm{MECC} := C\big(\bar{\mathcal{Q}};\mathcal G\big),~\mathrm{EMCC} := \sum_{s\in\mathcal S}p_sC(\mathcal{Q}_{s};\mathcal G),
\end{equation*}
where $\bar{\mathcal{Q}}:=\{\sum_{s\in\mathcal S}p_sQ_{se}|e\in\mathcal{E}\}$ and $\mathcal{Q}_s:=\{Q_{se}|e\in\mathcal{E}\}$.
In general, the MECC and the EMCC are not equal. Noting that the MCC is concave with respect to link capacities \cite{como13i}, one can show that the MECC is never less than the EMCC by Jensen's inequality. For instance, consider the numerical example in Section~\ref{sub_exm1}; it is shown that the EMCC equals 0.75 and MECC equals 1 in Fig.~\ref{fig_cut}.

\subsection{Main results}
The following investigates in which case there exist controls that attain the EMCC/MECC. We start from the simplest open-loop control. It is shown that if every link $e\in\mathcal{E}$ has infinite storage space without restricted receiving flows, there exists an open-loop control $\mu^{\mathrm{ol}}$ that guarantees the MECC. The control design is inspired by the solution of the maximum flow problem $(\mathrm{P}_1)$. Before presenting it, we denote by $u^*(\bar{\mathcal{Q}}):=\{u^*_{ej}(\bar{\mathcal{Q}})|(e,j)\in\mathcal{P}\}$ the solution of $(\mathrm{P}_1)$ given the set of link capacities $\bar{\mathcal{Q}}$ and define
\begin{equation}
    \bar{\gamma}_{ej} := \frac{u^*_{ej}({\bar{\mathcal{Q}}})}{\sum_{j'\in\mathcal{E}_e^+} u^*_{ej'}(\bar{\mathcal{Q}})},~(e,j)\in\mathcal{P}. \label{eq_gamma}
\end{equation}
Clearly, $\bar\gamma_{ej}$ denotes the proportion of the outflow of link $e$ routed to link $j$ per the solution $u^*(\bar{\mathcal{Q}})$.

Then the open-loop control is given by
\begin{equation}
    \mu_{ej}^{\mathrm{ol}} = \bar{\gamma}_{ej}Q_e^{\mathrm{max}}, ~(e,j)\in\mathcal{P},
    \label{eq_ol}
\end{equation}
where $Q_e^\max:=\max_{s\in\mathcal{S}}Q_{se}$. The formal result of the open-loop control is stated below:
\begin{thm}[Max-flow min-expected-cut]
\label{thm_infinite}
Consider an acyclic dynamic flow network with $x_e^{\mathrm{max}}=\infty$ and $r_e(s,x_e)=\infty$ for any link $e\in\mathcal{E}$. There exists an open-loop control $\mu^{\mathrm{ol}}$ given by \eqref{eq_ol} such that the network throughput equals the min-expected-cut capacity.
\end{thm}


%
However, few links in practice have infinite storage, which indicates the MECC is generally unattainable. For networks possibly with finite storage space, we consider mode-dependent control given observable disruption mode $S(t)$. Suppose a network with the following physical disruptions:
\begin{subequations}
\begin{align}
     f_e(s,x_e)&=\min\{f_e(x_e),F_{se}\}, 
     \label{eq_thm3_con1} 
     \\
     r_e(s,x_e)&=\min\{r_e(x_e),R_{se}\}.
     \label{eq_thm3_con2}
\end{align}
\end{subequations}
These disruptions restrict maximal outflows or inflows, which satisfy Assumption~\ref{asm_fault}.2. For the sake of simplicity of analysis, we also assume the critical density $x_e^c$ given by \eqref{eq_x_e_c} is finite for every $e\in\mathcal{E}$. If any disruption mode $s\in\mathcal{S}$ can be correctly observed, we design a mode-dependent control as follows:
\begin{align}
    \mu_{ej}^{\mathrm{md}}(s) := u_{ej}^*(\mathcal{Q}_{s}),~(e,j)\in\mathcal{P},
    \label{eq_mumd}
\end{align}
where $ u_{ej}^*(\mathcal{Q}_{s})$ is an optimal solution to the maximum flow problem $(\mathrm{P}_1)$ with the set of link capacities $\mathcal{Q}_s$. Then we state the following theorem:
\begin{thm}[Max-flow expected-min-cut]
\label{thm_observable}
Consider an acyclic dynamic flow network with a set of observable disruptions \eqref{eq_thm3_con1}-\eqref{eq_thm3_con2}. There exists a mode-dependent control $\mu^{\mathrm{md}}$ given by \eqref{eq_mumd} such that the network throughput equals the expected-min-cut capacity.
\end{thm}


\begin{table*}[h]
    \centering
    \scriptsize
    \caption{Comparison of resiliency scores of different control laws.}
    \resizebox{\textwidth}{!}{
    \begin{tabu}to\linewidth{X[0.5,c,m]X[0.6,c,m]X[0.6,c,m]X[0.2, c,m]X[0.2,c,m]X[0.2,c,m]||X[0.2,c,m]X[0.2,c,m]X[0.2,c,m]||X[0.2,c,m]X[0.2,c,m]X[0.2,c,m]}
        \toprule
        \multicolumn{3}{c}{} &  \multicolumn{3}{c}{Open-loop} &
        \multicolumn{3}{c}{Mode-dependent} &
        \multicolumn{3}{c}{Density-dependent} \\
        \midrule 
        Infinite space & { Cyber disruptions} & { Physical disruptions} & $\underline{\eta}^{\mathrm{ol}}_1$ &  $\underline{\eta}^{\mathrm{ol}}_2$ & 
        $\eta^{\mathrm{ol}}_{\mathrm{sim}}$ &
        $\underline{\eta}^{\mathrm{md}}_1$ &  $\underline{\eta}^{\mathrm{md}}_2$ & 
        $\eta^{\mathrm{md}}_{\mathrm{sim}}$ &
        $\underline{\eta}^{\mathrm{dd}}_1$ &  $\underline{\eta}^{\mathrm{dd}}_2$ & 
        $\eta^{\mathrm{dd}}_{\mathrm{sim}}$
        \\
        \midrule
        yes & no & no   & 1 & 1 & 1 & 1 & 1 & 1 & 1 & 1 & 1 \\
        yes & yes & no  & 1 & 1 & 1 & 1 & 1 & 1 & 1 & 1 & 1 \\
        yes & no & yes  & 1 & 1 & 1 & 0.750 & 0.750 & 0.750 & 1 & 1 & 1 \\
        yes & yes & yes & 1 & 1 & 1 & 0.750 & 0.750 & 0.750 & 1 & 1 & 1 \\
        no & no & no    & 1 & 1 & 1 & 1 & 1 & 1 & 1 & 1 & 1 \\
        no & yes & no   & 1 & 1 & 1 & 1 & 1 & 1 & 1 & 1 & 1 \\
        no & no & yes   & 0.747 & 0.850 & 0.859 & 0.750 & 0.750 & 0.750 & 0.746 & 0.867 & 0.873 \\
        no & yes & yes  & 0.747 & 0.850 & 0.859 & 0.750 & 0.750 & 0.750 & 0.739 & 0.851 & 0.863 \\
        \bottomrule
    \end{tabu}}
    \label{tab_control}
\end{table*}

Note that EMCC is less than the MECC in general. Hence the mode-dependent control above may appear conservative, provided a gap between the EMCC and the MECC. Besides, the mode-dependent control requires quick detection of disruptions and informs local controllers of disruption occurrences. If some of controllers are unaware of changes of disruption modes, Theorem~4 does not hold and the EMCC may not be guaranteed. In that case, we can apply Theorems~1 and 2 to the throughput analysis by modeling actuation faults.

By contrast, it is more practical to measure the network state $X(t)$. Thus we consider density-dependent control using the network state feedback. For a general network (with possibly incorrect sensing or with finite link storage spaces), one can apply the results of our resiliency analysis to design closed-loop controls.
Suppose the control law is parameterized by $\theta$. Then the parameter $\theta$ is designed by solving the following programming problem:
\begin{equation*}
    (\mathrm{P}_3)~\max_{\alpha\geq0,\theta\in\Theta}\alpha \quad s.t.~\eqref{eq_psG}.
\end{equation*}
Let $\theta^*$ and $\alpha^*$ denote optimal solutions of the programming problem $\mathrm{P}_3$. Then it follows from Theorem~\ref{thm_stable} that the throughput under the control law parameterized by $\theta^*$ is lower bounded by $\alpha^*$:
\begin{cor}[Throughput-guaranteed control]
\label{thm_general} 
Consider an acyclic dynamic flow network with a density-dependent control $\mu^{\mathrm{dd}}$ deriving from $\mathrm{P}_3$. Then, $\mu^{\mathrm{dd}}$ can attain a throughput lower-bounded by the optimal $\alpha^*$.
\end{cor}

To facilitate the control design, we can restrict the control structure $\theta\in\Theta$ so that $\mathcal{E}_{\mathrm{inf}}$ and $\{\mathcal{B}^+_{e}|e\in\mathcal{E}_{\mathrm{inf}}\}$ are fixed. In practice, the determination of these sets may depend on infrastructure capability or control preferences, e.g., whether link $e$ has sufficient storage space, or whether link $e$ is allowed to be a bottleneck.

The rest of this section first illustrates the above results with a numerical example (Section~\ref{sec_num_control}) and then proves them respectively (Sections~\ref{sec_pf_thm3_suf}-\ref{sec_pf_thm4}).

\subsection{Numerical example}
\label{sec_num_control}

Again, consider the network in Fig.~\ref{fig_network}. We use Theorems 3, 4 and Corollary 1 to design open-loop control $\mu^{\mathrm{ol}}$, mode-dependent control $\mu^{\mathrm{md}}$ and density-dependent control $\mu^{\mathrm{dd}}$, respectively. Table~\ref{tab_control} lists the resiliency scores in various scenarios. Note the nominal MCC equals one. Our discussions are as follows:
\begin{enumerate}[(i)]
    \item It is demonstrated that the open-loop control can attain the MECC when the links have infinite storage space. Besides, the results also show that the MECC is not guaranteed when the link storage space becomes finite. 
    \item It is verified that the mode-dependent control can attain the EMCC. Although this control seems conservative, it is still able to outperform the logit routing in case of links with infinite storage space; see Table~\ref{tab_analysis}.
    \item Compared with the open-loop control, density-dependent control can achieve higher resiliency scores when dealing with physical disruptions. However, it is vulnerable to cyber disruptions.
    \item The results indicate that we can improve network resiliency by designing control laws with higher lower bounds of the resiliency scores.
\end{enumerate}

Next, we present the design of three controls.
\subsubsection{Open-loop control}
\label{sec_ol}
Solving the maximum flow problem $(\mathrm{P}_1)$ with the expected link capacities gives $u_{e_oe_1}^*=u_{e_1e_3}^*=u_{e_oe_4}^*=u_{e_4e_5}^*=0.5$ and $u_{e_1e_2}^*=0$. According to \eqref{eq_ol}-\eqref{eq_gamma}, we have $\mu_{e_oe_1}^{\mathrm{ol}}=\mu_{e_1e_3}^{\mathrm{ol}}=\mu_{e_oe_4}^{\mathrm{ol}}=\mu_{e_4e_5}^{\mathrm{ol}}=0.5$ and $\mu_{e_1e_2}^{\mathrm{ol}}=0$.

\subsubsection{Mode-dependent control}
\label{sec_md}
The control $\mu^{\mathrm{md}}$ is obtained from solving the maximum flow problem $(\mathrm{P}_1)$ for each disruption mode $s$. Then we have $\mu_{e_oe_1}^{\mathrm{md}}=\mu_{e_1e_3}^{\mathrm{md}}=0.5$, $\mu_{e_1e_2}^{\mathrm{md}}=0$ and
\begin{equation*}
    \mu_{e_oe_4}^{\mathrm{md}}(s)=\mu_{e_4e_5}^{\mathrm{md}}(s) = \begin{cases}
    0.5, & \mathrm{if}~s\in\{s_0, s_2\},\\ 
    0,  & \mathrm{if}~s\in\{s_1, s_3\}.\\
    \end{cases}
\end{equation*}

\subsubsection{Density-dependent control}
\label{sec_dd}
We consider the density-dependent routing at node $v_o$, namely 
$\mu_{e_oe_1}^{\mathrm{dd}}(s, x) = w_c(u_1-x_1)$ and $\mu_{e_oe_4}^{\mathrm{dd}}(s, x) = w_c(u_4-T_{e_4}(s,x))$, where the gains are set as the congestion wave speed $w_c$, the parameters $u_1$ and $u_4$ are to be designed, and $T_{e_4}(s,x)$ is the fault mapping given in Table~\ref{tab_modes}. We interpret $u_1$ (resp. $u_4$) as the maximum density allowed by the control for link $e_1$ (resp. $e_4$). We also let $\mu_{e_1e_3}^{\mathrm{dd}}(s,x)=f_{e_1}(s,x_{e_1})$ and $\mu_{e_1e_2}^{\mathrm{dd}}(s,x)=0$.

When all links $e_o,e_1,\cdots,e_5$ have infinite storage space, solving $\mathrm{P}_3$ gives $u_1=u_4=\infty$. It means that the density-dependent control is reduced to an open-loop control that routes flows into links $e_1$ and $e_4$ with the same proportion. For links $e_1,\cdots,e_5$ only with finite storage space, our method yields $u_1=u_4=1.5$ for the network subject to either the cyber or physical disruption and  $u_1=2.3, u_4=1.5$ in case of both the cyber and physical disruptions. More details can be found in the supplementary material.

\subsection{Proof of Theorem~\ref{thm_infinite}}

\label{sec_pf_thm3_suf}
\subsubsection{Sufficiency}
The proof is similar to that of Theorem~\ref{thm_stable}. Noting $r_e(s,x_e)=\infty$ for any $e\in\mathcal{E}$, we obtain $\mathcal{B}_{e}^{+}=\varnothing$ for $e\in\mathcal{E}_{\mathrm{inf}}$. Consider the following Lyapunov function:
\begin{align*}
    V^{\mathrm{ol}}(s,x) :=& \sum_{e\in\mathcal{E}_{\mathrm{inf}}} a_e x_e[\frac{1}{2} x_e  + \sum_{m\in\mathcal{A}^{-}_{e}} \bar{\gamma}_{me} x_m + b_{se}],
\end{align*}
where $\bar{\gamma}_{me}$ is recursively defined by
\begin{equation}
    \bar{\gamma}_{me} = \sum_{i\in\mathcal{A}_m^+\cap\mathcal{E}_e^-} \bar{\gamma}_{mi}\bar{\gamma}_{ie} \label{eq_gamma_recursive}
\end{equation}
with $\bar{\gamma}_{ie}$ given by \eqref{eq_gamma}. Noting the physical meaning of $\bar{\gamma}_{ie}$, we interpret $\bar{\gamma}_{me}$ as the proportion of the outflow of link $m$ routed to link $e$ per the solution $u^*(\bar{\mathcal{Q}})$. Specially, for link $e_o$ with a demand $\alpha=C(\bar{\mathcal{Q}};\mathcal{G})$, we have
\begin{equation}
    \bar{\gamma}_{e_oe}C(\bar{\mathcal{Q}};\mathcal{G}) = \sum_{i\in\mathcal{E}_j^-}u_{ie}^*(\bar{\mathcal{Q}}). \label{eq_gamma_important}
\end{equation}
The above equation holds because, as specified by the solution $u^*(\bar{\mathcal{Q}})$, the optimal demand $\alpha^*$ equals $C(\bar{\mathcal{Q}};\mathcal{G})$ and the flow routed from link $e_o$ to link $e$ equals $\sum_{i\in\mathcal{E}_j^-}u_{ie}^*(\bar{\mathcal{Q}})$. 

The Lyapunov function indicates that we only need to show
\begin{equation}
    \sum_{s\in\mathcal S}p_s \max_{x\in\mathcal{D}^{\mathrm{ol}}_{e}} N_{e}(s,x) < 0,~\forall \alpha< C(\bar{\mathcal{Q}}; \mathcal{G}),\forall e\in\mathcal{E}_{\mathrm{inf}}, \label{eq_pf_thm4_1}
\end{equation}
where $\mathcal{D}_e^{\mathrm{ol}}=\{x\in\mathcal{X}|x_e\geq x_e^{c*}\}$ and
$N_e(s,x)=\sum_{m\in\mathcal{A}_e^-}\bar{\gamma}_{me}G_m(s,x)+G_e(s,x)$.

The open-loop control yields  $q_{mj}^{\mathrm{ol}}(s,x)=\bar{\gamma}_{mj}f_{m}(s,x_m)$. Thus, for any link $m\in\mathcal{A}_e^-$, we have 
\begin{align*}
    & - \sum_{j\in\mathcal{E}_m^+} \bar{\gamma}_{me}q_{mj}^{\mathrm{ol}}(s, x) 
    + \sum_{j'\in\mathcal{E}_m^+\cap\mathcal{A}_e^-} \bar{\gamma}_{j'e} q_{mj'}^{\mathrm{ol}}(s, x) \\
    =&-\sum_{j\in\mathcal{E}_m^+} \bar{\gamma}_{me}\bar{\gamma}_{mj}f_{m}(s, x_m)  +
    \sum_{j'\in\mathcal{E}_m^+\cap\mathcal{A}_e^-} \bar{\gamma}_{j'e}\bar{\gamma}_{mj'} f_{m}(s, x_m) \\
    \overset{\tiny \eqref{eq_gamma}}{=}& -\bar{\gamma}_{me}f_m(s,x_m) + \bar{\gamma}_{me}f_m(s,x_m) =  0.
\end{align*}
So we obtain $N_e(s,x)=\bar{\gamma}_{e_oe}\alpha-f_e(s,x_e)$ and
\begin{equation}
    \sum_{s\in\mathcal S}p_s \max_{x\in\mathcal{D}^{\mathrm{ol}}_{e}} N_{e}(s,x) = \bar{\gamma}_{e_oe}\alpha - \sum_{s\in\mathcal{S}} p_s Q_{se},~\forall e\in\mathcal{E}_{\mathrm{inf}}. \label{eq_pf_thm4_2}
\end{equation}
Recalling \eqref{eq_gamma_important}, we arrive at 
\begin{equation}
\bar{\gamma}_{e_oe}C(\bar{\mathcal{Q}};\mathcal{G})=\sum_{i\in\mathcal{E}_e^-} u^*_{ie}(\bar{\mathcal{Q}}) \leq \sum_{s\in\mathcal{S}}p_sQ_{se} \label{eq_pf_thm4_3}
\end{equation}
where the last inequality holds because of the constraint in the maximum flow problem. Combining \eqref{eq_pf_thm4_2} and \eqref{eq_pf_thm4_3}, we complete the proof of \eqref{eq_pf_thm4_1}. \qed

\subsubsection{Necessity}
We prove the necessity by definition. Consider the cut-set $\mathcal{C}$ with the min-cut capacity $C(\bar{\mathcal{Q}};\mathcal{G})$, where $\mathcal{C}$ is a set of links that divide the network into two disjoint parts containing the origin $v_o$ and the destination $v_d$ respectively. Let $\mathcal{E}_{\mathcal{C}}^-$ denote the set of links upstream of the cut-set. By the open-loop control, the outflow from $\mathcal{E}_{\mathcal{C}}^-\cup\mathcal{C}$ does not exceed $C(\bar{\mathcal{Q}};\mathcal{G})$. If the inflow $\alpha$ is not less than $C(\bar{\mathcal{Q}};\mathcal{G})$ and $Z<\infty$ is given, there must exist some initial condition such that \eqref{eq_bounded} fails to hold. It concludes the proof. \qed

\subsection{Proof of Theorem~\ref{thm_observable}}
\label{sec_pf_thm4}

\subsubsection{Sufficiency}
The proof is based on Theorem~1. Noting Lemma~\ref{lmm_md} proved in Appendix~\ref{app_lmm_md}, we only need to show 
\begin{equation}
    \sum_{s\in\mathcal S}p_s \max_{x\in\mathcal{D}^{\mathrm{md}}_{e_o}} (\alpha-\sum_{j\in\mathcal{E}_{e_o}^+}q_{e_oj}^{\mathrm{md}}(s,x)) < 0,~\forall \alpha<\sum_{s\in\mathcal{S}}p_s C(\mathcal{Q}_s; \mathcal{G}). \label{eq_pf_thm3_1}
\end{equation}
\begin{lmm} \label{lmm_md}
The mode-dependent control $\mu^{\mathrm{md}}$ admits an invariant set
\begin{equation*}
     \mathcal{X}^{\mathrm{md}} = [\underline{x}_{e_o}^{\mathrm{md}}, \infty)\times\prod\limits_{e\neq e_o}[\underline{x}_e^{\mathrm{md}},\bar{x}_e^{\mathrm{md}}] = [\alpha, \infty)\times\prod\limits_{e\neq e_o}[0,x_e^{c*}],
\end{equation*}
which leads to $\mathcal{E}_{\mathrm{inf}}=\{e_o\}$ and $\mathcal{B}_{e_o}^{+}=\varnothing$.
\end{lmm}

Then it follows
\begin{align*}
    \sum_{s\in\mathcal S}p_s \max_{x\in\mathcal{D}^{\mathrm{md}}_{e_o}} (\alpha-\sum_{j\in\mathcal{E}_{e_o}^+}q_{e_oj}^{\mathrm{md}}(s,x))
    =&\sum_{s\in\mathcal S}p_s (\alpha-\sum_{j\in\mathcal{E}_{e_o}^+} \mu_{e_oj}^{\mathrm{md}}(s)),
\end{align*}
where the equality is due to i) $\mathcal{B}_{e_o}^{+}=\varnothing$, ii) $f_{e_o}(s,x_{e_o})\geq Q_{se_o}$ for $x_{e_o}\geq x_{e_o}^{c*}$ and iii) the constraint \eqref{eq_p1_4}. By the constraint \eqref{eq_p1_1}, we obtain $\sum_{j\in\mathcal{E}_{e_o}^-}\mu_{e_oj}^{\mathrm{md}}(s) =\alpha^*(\mathcal{Q}_s; \mathcal{G}) = C(\mathcal{Q}_s; \mathcal{G})$, which completes the proof of \eqref{eq_pf_thm3_1}. \qed

\subsubsection{Necessity} We consider the set $\{e_o\}$. 
Then the remaining proof is similar to that for the necessity of Theorem~\ref{thm_infinite}. \qed


\section{Concluding remarks}
\label{sec_conclude}

This paper investigates resiliency and control design of dynamic flow networks suffering cyber-physical disruptions. First, we apply piecewise-deterministic Markov process to modeling disruptions and their impacts on networks. Then, a set of stability criteria are proposed with the physical insights into flow networks. We also simplify these criteria with the monotone network dynamics. The stability conditions enable the measurement of resiliency in terms of throughput. They also contribute to two specific resiliency-by-design controls. The first one attains the min-expected-cut capacity if every link has infinite storage space; the second one attains the expected-min-cut capacity if the disruptions decrease maximum sending and/or receiving flows. In the general case, we propose a density-dependent control with a lower-bounded throughput. Our numerical examples show that the stability condition can yield tight lower bounds of resiliency scores and  enhance resiliency by designing control laws that raise the lower bounds.


This work can serve as a basis for multiple future studies. First, our sufficient stability conditions only yield lower bounds of throughput. It is tempting to derive upper bounds by investigating the necessary conditions. Besides, this paper exploits the property of monotone flow networks to reduce the computation costs of verifying our stability conditions.  However, monotone dynamics could be unavailable in multi-commodity networks \cite{nilsson2014resilience}. It is worthwhile considering networks with weaker properties, such as mixed monotonicity \cite{como2017resilient}. Finally, this paper formulates the control design problem for the whole network. Solving it for large-scale networks could not be easy. Thus how to design control laws with guaranteed performances but only using a part of links/nodes is worth investigating. 
\begin{appendix}
\section{Appendices}

\subsection{Open-loop control in the nominal case}
\label{app_olnominal}

We consider an open-loop control policy as follows:
\begin{equation}
    \hat{\mu}^{\mathrm{ol}}_{ej} = \hat{\gamma}_{ej}Q_{s_0e}, \label{eq_naive_ol}
\end{equation}
where
\begin{equation}
    \hat{\gamma}_{ej} := \frac{u^*_{ej}({{\mathcal{Q}}}_{s_0})}{\sum_{j'\in\mathcal{E}_e^+} u^*_{ej'}({\mathcal{Q}}_{s_0})},~(e,j)\in\mathcal{P}. \label{eq_gamma_1}
\end{equation}
Clearly, it has the same structure as the open-loop control $\mu^{\mathrm{ol}}_{ej}$ given by \eqref{eq_ol}.

Before we state the lemma, we define \begin{equation}
    \hat{\gamma}_{me} = \sum_{i\in\mathcal{A}_m^+\cap\mathcal{E}_e^-} \hat{\gamma}_{mi}\hat{\gamma}_{ie}. \label{eq_gamma_2}
\end{equation}
Noting \eqref{eq_gamma_recursive} and \eqref{eq_pf_thm4_3}, for any $\alpha<C({\mathcal{Q}}_{s_0};\mathcal{G})$, we obtain in a similar way
 \begin{equation}
    \hat{\gamma}_{e_oe}\alpha<\hat{\gamma}_{e_oe}C({\mathcal{Q}}_{s_0};\mathcal{G}) = \sum_{i\in\mathcal{E}_j^-}u_{ie}^*({\mathcal{Q}}_{s_0}) \leq Q_{s_0e}. \label{eq_gamma_important_2}
\end{equation}

\begin{lmm}
\label{lmm_ol_nominal}
Consider an acyclic network satisfying Assumption~3 and only admitting the nominal mode $s_0$. Given any demand $\alpha<C(\mathcal{Q}_{s_0};\mathcal{G})$ and the open-loop control given by \eqref{eq_naive_ol}, there exists a stable equilibrium such that $f_e(s_0,\hat{x}_e)=\hat{\gamma}_{e_oe}\alpha$ for any $e\in\mathcal{E}$. Moreover, if there exists a neighborhood $U(\hat{x})$ of $\hat{x}$ such that $f_{e}(s_0,x_e)$ is strictly increasing in $x_e$ for every $e\in\mathcal{E}$, the equilibrium $\hat{x}$ is also globally asymptotically stable.
\end{lmm}

Obviously, the global asymptotic stability stated in Lemma~\ref{lmm_ol_nominal} implies that the open-loop control \eqref{eq_naive_ol} can enable the throughput to achieve the nominal min-cut capacity. Note that the required condition is really mild. As we will see in the proof, all of the link densities are fewer than their corresponding critical densities. It means that none of the links are congested. In that case, the sending flow $f_e(s_0,x_e)$ is typically sensitive to density changes, i.e., strictly increasing in $x_e$.

Below we give the proof of Lemma~\ref{lmm_ol_nominal}.
\begin{proof}
We first show the existence of $\hat{x}$. For any demand $\alpha<C(\mathcal{Q}_{s_0};\mathcal{G})$, we conclude $\hat{x}\in\prod_{e\in\mathcal{E}}[0,x_{s_0e}^c]$ by using Assumptions~3.1-3.2 and $f_e(s_0,\hat{x}_e)=\hat{\gamma}_{e_oe}\alpha < Q_{s_0e}$, where $x_{s_0e}^c$ is given by \eqref{eq_mode_crti}. 

Next, we prove that $\hat{x}$ is a stable equilibrium. Again noting \eqref{eq_naive_ol} and \eqref{eq_gamma_important_2}, we conclude that there exists a neighbourhood $W(\hat{x})$ of $\hat{x}$ such that
$q_{ej}(s_0, x) = \hat{\gamma}_{ej}f_e(s_0,x_e)$.
Then $\hat{x}$ is an equilibrium by \eqref{eq_gamma_2}.
The stability of $\hat{x}$ is directly implied by Theorem 6 (iii) in \cite{como2017resilient}.

Note that $\hat{x}$ is globally asymptotically stable if and only if it is locally asymptotically stable; see Theorem 6 (iv) in \cite{como2017resilient}. Below we show that the network is locally asymptotically stable over the neighborhood  $W(\hat{x})\cap U(\hat{x})$. We first notice link $x_{e_o}$ converges to $\hat{x}_{e_o}$ by using the monotonicity and Lipschtiz continuity of $f_{e_o}(s_0,x_{e_o})$. Then we consider links from the upstream to the downstream. We can show $x_e$ converges by using the convergence of its upstream links. 
\end{proof}

\subsection{Extension for cyclic networks}
\label{app_cyclic}
We present stability analysis for cyclic networks. One major difference between cyclic and acyclic networks is $\mathcal{A}_e^-\cap\mathcal{A}_e^+\neq\varnothing$ for some $e\in\mathcal{E}$ in a cyclic network. It induces that Lemma~\ref{lmm_ak} may not hold because of $\mathcal{A}_{e,\mathrm{inf}}^{-}\cap\mathcal{A}_{e,\mathrm{inf}}^{+}\neq\varnothing$. We address this problem by considering a partition of $\mathcal{E}_{\mathrm{inf}}$.

We denote by $\{\mathcal{L}_{z}\}_{z=1}^Z$ a partition of $\mathcal{E}_{\mathrm{inf}}$ such that given any $z_1,z_2\in\{1,2,\cdots,Z\}$ with $z_1\neq z_2$,
\begin{subequations}
\begin{align}
    e\leftrightarrow j,&~\forall e,j\in\mathcal{L}_{z_1}, e\neq j, \label{eq_partition1} \\
    e\not\leftrightarrow j,&~\forall e\in\mathcal{L}_{z_1},j\in\mathcal{L}_{z_2}, \label{eq_partition2}
\end{align}
\end{subequations}
where $e\leftrightarrow j$ represents that the two links are reachable from each other, and $e\not\leftrightarrow j$ denotes that at least one link is unreachable from the other. 
For a cyclic network, a partition can be obtained by merging the sets above whose elements are accessible from each other, and we can prove by contradiction the uniqueness of the partition.

Then we show by the partition $\{\mathcal{L}_{z}\}_{z=1}^Z$ that Theorem~\ref{thm_stable} holds for acyclic networks. It implies that the other stability results in Section III also apply to acyclic networks.
\begin{prp}
Consider a cyclic network satisfying Assumptions~\ref{asm_fault}.1-\ref{asm_fault}.3. Suppose that the network admits a demand $\alpha$ and an invariant set $\mathcal X^\mu\subseteq\prod_{e\in\mathcal{E}}[\underline{x}_e^\mu, \bar{x}_e^\mu]$ under a control law $\mu:\mathcal S\times\mathcal X\to\mathbb R_{\ge0}^{\mathcal{P}}$. The network is stable if \eqref{eq_psG} holds.
\end{prp}
\begin{proof}
Consider the following Lyapunov function:
\begin{align}
    \hat{V}(s,x) :=& \sum_{e\in\mathcal{E}_{\mathrm{inf}}} a_{\iota_e} x_{e}\Big[\frac{1}{2}\sum_{i\in\mathcal{L}_{\iota_e}} x_i +\sum_{m\in\mathcal{A}_{e}^{-}\setminus\mathcal{L}_{\iota_e}} x_m  + \sum_{\ell\in\mathcal{B}_{e}^+}\int^{x_{\ell}}_{\underline{x}_{\ell}}\rho_{\ell}(\zeta)\mathrm{d}\zeta + b_{se}\Big], \label{eq_Ly3}
\end{align}
where $\iota_e$ denotes the set index such that $e\in\mathcal{L}_{\iota_e}$. The rest of the proof is similar to that of Theorem~\ref{thm_stable}, except Lemma~\ref{lmm_ak}. We instead use the following lemma:
\begin{lmm}\label{lmm_ak_2}
The system of equations
\begin{align}
    \pi_{e} a_{\iota_e} + \bar{G} \sum_{n\in\mathcal{A}_{e,\mathrm{inf}}^{+}\setminus\mathcal{L}_{\iota_e}} a_{\iota_n} < 0, ~\forall e\in{\mathcal E}_{\mathrm{inf}}
    \label{eq_ak_2}
\end{align}
has a positive solution for $\{a_{\iota_e}|e\in\mathcal{E}_{\mathrm{inf}}\}$ if $\pi_e<0$ for any $e\in\mathcal{E}_{\mathrm{inf}}$.
\end{lmm}
Note that \eqref{eq_partition1}-\eqref{eq_partition2} reveals an acyclic structure of the partition. Thus we can prove Lemma~\ref{lmm_ak_2} in a similar way to Lemma~\ref{lmm_ak}.
\end{proof}

\subsection{Proof of Lemma~\ref{lmm_1}}
\label{app_lmm1}
To show the existence of a solution, note that \eqref{eq_bks} is equivalent to $\Lambda y = p^{\mathrm{T}}z-z$ where $z=[z_{s_0},\cdots,z_{s_M}]^{\mathrm{T}}$. Since the discrete state process is ergodic, the rank of the matrix $\Lambda$ is $M$.
Scaling row $i+1$ with $p_{s_i}$ for $i=0,1,\cdots,M$ and adding the scaled rows $1,2,\cdots,M$ to row $M+1$, we obtain
\begin{equation}
\label{augmented}
\left[\begin{array}{ccc}
    -p_{s_0}\sum\limits_{i\neq0}\lambda_{s_0s_i} & \cdots & p_{s_0}\lambda_{{s_0}{s_M}} \\
        \vdots & \ddots & \vdots\\
        \sum\limits_{i\neq0}p_{s_i}\lambda_{s_is_0}-p_{s_0}\sum\limits_{i\neq0}\lambda_{s_0s_i} & \cdots & \sum\limits_{i\neq m}p_{s_i}\lambda_{s_is_M}-p_{s_M}\sum\limits_{i\neq M}\lambda_{s_Ms_i}
    \end{array}\right]y =\tilde{z}
\end{equation}
where $\tilde{z}=[p_{s_0}(p^{\mathrm{T}}z-z_{s_0}),\cdots,
(\sum_{i=0}^Mp_{s_i})p^{\mathrm{T}}z-p^{\mathrm{T}}z]^{\mathrm{T}}$. Noting $\Lambda^{\mathrm{T}}p = \bm{0}$, we know that
$\sum_{i\neq0}p_{s_i}\lambda_{s_is_j}-p_{s_j}\sum_{i\neq j}\lambda_{s_js_i}=0$ for any $j=0,\cdots,M$. Also note that
$(\sum_{i=0}^Mp_{s_i})p^{\mathrm{T}}z-p^{\mathrm{T}}z=0$.
Hence, the rank of the augmented coefficient matrix of the system of linear equations \eqref{augmented} is also $M$, equal to the rank of the coefficient matrix. Therefore, \eqref{augmented} must have a particular solution $y^{\mathrm{p}}$. Noting that $\mathrm{rank}(\Lambda)=M$ and that $y=\omega\bm{1}$ is a solution to $\Lambda y = \bm{0}$ for any $\omega\in\mathbb{R}$, we can conclude that the general solution of \eqref{eq_bks} is given by $y^{\mathrm{g}}=y^{\mathrm{p}}+\omega\bm{1}$. Clearly, a non-negative solution is available by letting $\omega$ be a sufficiently large number.  \qed


\subsection{Proof of Lemma~\ref{lmm_ak}}
\label{app_lmm2}
If $\bar{G}\le0$, the lemma is trivial. Now we consider $\bar{G}>0$ and construct $\{a_e|e\in\mathcal{E}_{\mathrm{inf}}\}$ satisfying \eqref{eq_ak}. Let $\bar{\pi}:=\max_{e\in\mathcal{E}_{\mathrm{inf}}}\pi_e<0$ and $K:=|\mathcal{E}_{\mathrm{inf}}|$. Since this lemma focuses on acyclic networks, we can iteratively consider the links in $\mathcal{E}_{\mathrm{inf}}$ backwards in some sense.


We first assume $\bar{G} \ge -\bar{\pi}$. The construction starts with initializing $\mathcal{N}^{(0)}:=\mathcal{E}_{\mathrm{inf}}$ and $i:=1$. In the $i$-th step, we consider 
$\mathcal{M}^{(i-1)}:=\{e\in\mathcal{N}^{(i-1)}|\forall i\in\mathcal{N}^{(i-1)}\setminus\{e\},i\not\to e\}$, where $i\not\to e$ implies that there are no directed paths from link $i$ to link $e$. Intuitively, $\mathcal{M}^{(i-1)}$ denotes the collection of the far downstream links among $\mathcal{N}^{(i-1)}$. For any $e\in\mathcal{M}^{(i-1)}$, we let $a_e = (-\bar{G}K/\bar{\pi})^{i-1}>0$ and then obtain
\begin{align*}
    &\pi_{e} a_{e} + \bar{G} \sum\limits_{n\in \mathcal{A}_{e,\mathrm{inf}}^{+}} a_{n} <  \bar{\pi} 
    (-\frac{\bar{G}K}{\bar{\pi}})^{i-1} + \bar{G}K (-\frac{\bar{G}K}{\bar{\pi}})^{i-2}  =  0.
\end{align*}
The $i$-th step ends up with $\mathcal{N}^{(i)}=\mathcal{N}^{(i-1)}\setminus\mathcal{M}^{(i-1)}$. The iteration stops when $\mathcal{N}^{(i)}=\varnothing$, which indicates that all the links of $\mathcal{E}_{\mathrm{inf}}$ are considered.

If $\bar{G} < -\bar{\pi}$, the same iteration is conducted except letting $a_e = -(\bar{G} K^{i-1})/\bar{\pi}>0$ in the $i$-th step. \qed


\subsection{Proof of Lemma~\ref{lmm_md}}
\label{app_lmm_md}

First, we note that the disruptions \eqref{eq_thm3_con1}-\eqref{eq_thm3_con2} imply 
\begin{subequations}
\begin{align}
    Q_{se} &\leq r_e(s,x_e),~\forall s\in\mathcal{S},\forall x_e\leq x_{e}^{c*}. \label{eq_pf_lmm4_2} \\
    Q_{se} &\leq f_e(s,x_e),~\forall s\in\mathcal{S},\forall x_e\geq x_{e}^{c*}, \label{eq_pf_lmm4_3}
\end{align}
\end{subequations}
where $x_e^{c*}$ is given by \eqref{eq_x_e_c}.
Besides, for any $x\in\mathcal{X}$ with $x_j\leq x_j^{c*}$, we obtain
\begin{equation}
    r_{ej}^{\mathrm{md}}(s,x) = \frac{\mu_{ej}^{\mathrm{md}}(s)}{\sum_{i\in\mathcal{E}_j^-}\mu_{ij}^{\mathrm{md}}(s)}r_j(s,x_j) \overset{\tiny\eqref{eq_p1_4},\eqref{eq_pf_lmm4_2}}{\geq} \mu_{ej}^{\mathrm{md}}(s). \label{eq_pf_lmm4_4}
\end{equation}

To show that $\mathcal{X}^{\mathrm{md}}$ is an positively invariant set, we prove that on the boundary of $\mathcal{X}^{\mathrm{md}}$, the vector field $G(s,x)$ points towards the interior regardless of $s\in\mathcal{S}$. That is, for any $e\neq e_o\in\mathcal{E}$ and $(s,x)\in\mathcal{S}\times\mathcal{X}^{\mathrm{md}}$ with $x_e=x_e^{c*}$, we have
\begin{align*}
    G_e(s, x)
    \overset{\tiny \eqref{eq_q},\eqref{eq_pf_lmm4_4}}&{\leq} \sum_{i\in\mathcal{E}_e^-}\mu_{ie}^{\mathrm{md}}(s) -  \sum_{j\in\mathcal{E}_e^+}\min\{\mu_{ej}^{\mathrm{md}}(s),f_{ej}^{\mathrm{md}}(s,x)\} \\
    %
    %
    \overset{\tiny \eqref{eq_f_ej},\eqref{eq_pf_lmm4_3}}&{\leq} \sum_{i\in\mathcal{E}_e^-}\mu_{ie}^{\mathrm{md}}(s) - \min\{ \sum_{j\in\mathcal{E}_e^+}\mu_{ej}^{\mathrm{md}}(s),Q_{se}\} \\
    \overset{\tiny \eqref{eq_p1_2},\eqref{eq_p1_4}}&{=} 0.
\end{align*}

To show that $\mathcal{X}^{\mathrm{md}}$ is global attracting, we perform an induction on links. We first consider link $e$ reaching the destination, i.e. $\tau_e=v_{d}$, since it does not have downstream bottlenecks. We see $G_e(s,x)\leq 0$ for any $(s,x)\in\{(s,x)\in\mathcal{S}\times\mathcal{X}|x_e>x_e^{c*}\}$. In addition, given $\alpha<\sum_{s\in\mathcal{S}}p_s C(\mathcal{Q}_s; \mathcal{G})$, there exists some mode $s\in\mathcal{S}$ under which $x_e$ converges to a value strictly less than $x_e^{c*}$. Since the process $\{S(t);t\geq0\}$ is ergodic, it is concluded that $x_e$ enters $[0,x_e^{c*}]$ almost surely. Then, as indicated by \eqref{eq_pf_lmm4_4}, link $e$ is not a bottleneck for its upstream links under the mode-dependent control. The similar proof can be iteratively applied to links upstream of link $e$ until all links are considered.

The invariant set $\mathcal{X}^{\mathrm{md}}$ immediately gives $\mathcal{E}_{\mathrm{inf}}=\{e_o\}$; by noting \eqref{eq_pf_lmm4_4}, we also obtain $\mathcal{B}_{e_o}^{+}=\varnothing$. \qed




\end{appendix}


\bibliographystyle{IEEEtran}
\bibliography{Bibliography}



\end{document}